\documentclass{amia}
\usepackage{lipsum} 
\usepackage{multirow}
\usepackage{amssymb}
\usepackage{amsmath}
\usepackage{algorithm}
\usepackage{algpseudocode}
\usepackage{booktabs}
\usepackage{xcolor}
\usepackage{longtable,natbib}
\usepackage{wrapfig}
\usepackage{float}
\usepackage{makecell}
\usepackage{enumitem}
\usepackage{booktabs}
\usepackage{subfigure}
\usepackage{hyperref}

\usepackage{tikz}

\def\Algnameunderline{\underline{Ra}re \underline{Dis}ease \underline{Q}uestion-and-\underline{A}nswer}
\def\Algnameabbr{ReDis-QA}

\def\corpusname{ReCOP}

\definecolor{lightred}{RGB}{247, 220, 220}
\definecolor{lightgreen}{RGB}{220, 247, 227}
\definecolor{darkred}{HTML}{be002f}
\definecolor{darkgreen}{HTML}{00a64f}

\begin{document}

\title{Assessing and Enhancing Large Language Models in \\ Rare Disease Question-answering}

\author{Guanchu Wang$^*$$\dagger$, M.S., Junhao Ran$^*$$\dagger$, B.Eng., Ruixiang Tang$\dagger$, Ph.D., \\ Chia-Yuan Chang$\S$, M.S., Yu-Neng Chuang$\dagger$, M.S., Zirui Liu$\dagger$, Ph.D., \\ Vladimir Braverman$\dagger$, Ph.D., Zhandong Liu$\ddagger$, Ph.D., Xia Hu$\dagger$, Ph.D.}
\institutes{
    $\dagger$ Rice University;
    $\ddagger$ Baylor College of Medicine;
    $\S$ Texas A\&M University
}

\renewcommand{\thefootnote}{\fnsymbol{footnote}}
\footnotetext[0]{\!\!\!\! * Equal contribution.}
\renewcommand{\thefootnote}{\arabic{footnote}}

\maketitle

\section*{Abstract}

\textit{Despite the impressive capabilities of Large Language Models~(LLMs) in general medical domains, questions remain about their performance in diagnosing rare diseases.
To answer this question, we aim to assess the diagnostic performance of LLMs in rare diseases, and explore methods to enhance their effectiveness in this area.
In this work, we introduce a rare disease question-answering~(\Algnameabbr{}) dataset to evaluate the performance LLMs in diagnosing rare diseases. 
Specifically, we collected 1360 high-quality question-answer pairs within the \Algnameabbr{} dataset, covering 205 rare diseases. 
Additionally, we annotated meta-data for each question, facilitating the extraction of subsets specific to any given disease and its property. 
Based on the \Algnameabbr{} dataset, we benchmarked several open-source LLMs, revealing that diagnosing rare diseases remains a significant challenge for these models.
}

\textit{To facilitate retrieval augmentation generation for rare disease diagnosis, we collect the first rare diseases corpus (\corpusname{}), sourced from the National Organization for Rare Disorders (NORD) database. 
Specifically, we split the report of each rare disease into multiple chunks, each representing a different property of the disease, including their overview, symptoms, causes, effects, related disorders, diagnosis, and standard therapies. This structure ensures that the information within each chunk aligns consistently with a question.
Experiment results demonstrate that \corpusname{} can effectively improve the accuracy of LLMs on the \Algnameabbr{} dataset by an average of $\ 8\%$.
Moreover, it significantly guilds LLMs to generate trustworthy answers and explanations that can be traced back to existing literature.
The \Algnameabbr{} dataset, \corpusname{} corpus, and source codes of benchmark experiments are open-sourced at here\footnote{\Algnameabbr{} is open-sourced at \href{https://huggingface.co/datasets/guan-wang/ReDis-QA}{\textcolor{blue}{https://huggingface.co/datasets/guan-wang/ReDis-QA}}.}
\footnote{
\corpusname{} is open-sourced at \href{https://huggingface.co/datasets/guan-wang/ReCOP}{\textcolor{blue}{https://huggingface.co/datasets/guan-wang/ReCOP}}.}
\footnote{Source codes of benchmark experiments are open-sourced at \href{https://github.com/guanchuwang/redis-bench}{\textcolor{blue}{https://github.com/guanchuwang/redis-bench}}.
}
}


%

\section{Introduction}

Large language models~(LLMs) show magnificent power in natural language processing, causal inference, code generation, and have widely applied to medical research~\cite{clusmann2023future, thirunavukarasu2023large}.
The flexibility and general capacity of LLMs allow them to be easily deployed to different scenarios of medical research, such as patient trail matching~\cite{yuan2023large}, healthcare chatbot~\cite{yang2023large, bhayana2024chatbots}, and clinical note summarization~\cite{chuang2024spec}.
For example, LLMs like Llama~\cite{touvron2023llama}, Mistral~\cite{jiang2024mixtral}, Phi~\cite{abdin2024phi}, Gemma~\cite{team2024gemma} can work on medical tasks in a zero-shot manner based on simple medical-related instructions.
Additionally, they can be further enhanced with adaptations or alignments on medical corpora, such as the PMC-Llama~\cite{wu2024pmc}, Me-Llama~\cite{xie2024me}, and BioMistral~\cite{labrak2024biomistral}.
Despite the impressive capabilities of LLMs in general medical domains, \emph{questions persist about how LLMs perform in diagnosing rare diseases?}

%


Although rare diseases affect only a small portion of the population, they collectively impose substantial burdens on public health, affecting millions of individuals worldwide~\cite{nguengang2020estimating, mao2024ai}.
Diagnosing and treating these conditions are particularly challenging due to their complex genetic origins and unpredictable clinical manifestations~\cite{stark2023genomic, kafkas2023application}, requiring significantly intellectual decisions based on a vast knowledge base.
LLMs are pre-trained on the corpora comprising trillions of tokens, rivaling the expertise of human specialists~\cite{yu2023leveraging, brown2020language}. 
This extensive training enables LLMs to acquire sufficient medical knowledge for diagnosing common diseases~\cite{jahan2024comprehensive, nori2023capabilities, nori2023can}. 
However, LLMs still encounter several challenges when addressing tasks related to rare diseases, including:
\begin{itemize}


    
    \item \textbf{Generalization Challenges of LLMs to Rare Diseases.}
    Rare diseases affect a small number of people and are documented in limited literature. Pre-trained LLMs often struggle to generalize to rare diseases due to the unique and varied manifestations of these conditions. Consequently, LLMs might hallucinate by incorrectly linking rare diseases with common ones, leading to misleading results when queried about rare diseases.

    
    \item \textbf{Genetic Causes of Rare Diseases.} Many rare diseases are caused by genetic mutations, which requires detailed genetic information and understanding that may not be captured fully by LLMs pre-trained on general data.
    Moreover, the shortfall in genetic data cannot be effectively addressed through existing prompting strategies, highlighting a critical gap in their capabilities.

    
\end{itemize}
To answer the question about the capabilities of LLMs in diagnosing rare diseases, we aim to assess the capabilities of LLMs in rare disease diagnosis, exploring tailored datasets that could bridge the gap in their current performance.

\begin{figure}[t]
    \centering
    \includegraphics[width=0.95\linewidth]{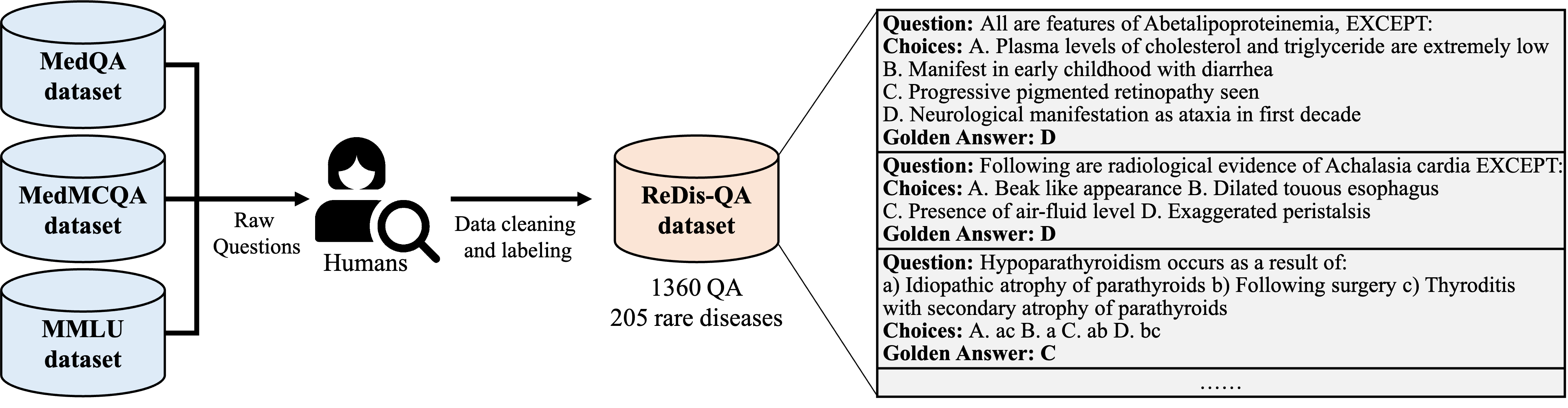}
    \caption{Pipeline of building the rare disease QA (\Algnameabbr{}) dataset: data collection, cleaning, and labeling.}
    \label{fig:dataset-collection} 
\end{figure}












    

In this work, we introduce a rare disease question-answering~(\Algnameabbr{}) dataset to evaluate the performance LLMs in diagnosing rare diseases. 
Specifically, we collect 1360 high-quality question-answer pairs within the \Algnameabbr{} dataset, spanning 205 rare diseases. 
Additionally, we annotated meta-data for each question, facilitating the extraction of subsets specific to any given disease and its property. 
Based on the \Algnameabbr{} dataset, we benchmark several open-source LLMs, revealing that diagnosing rare diseases remains a formidable challenge for current open-source LLMs.
To improve LLMs' performance, we collect the first rare diseases corpus (\corpusname{}), sourced from the National Organization for Rare Disorders (NORD) database. 
This database provides reliable and comprehensive reports on known rare diseases. 
The primary objective of \corpusname{} is to enhance the diagnostic capabilities of LLMs for rare diseases through retrieval augmented generation (RAG).
To better fit the RAG framework, we split the report of each rare disease into multiple chunks, each representing a different property of the disease, including overview, symptoms, causes, effects, related disorders, diagnosis, and standard therapies. This structure ensures that the information within each chunk aligns consistently with a question.
Experiment results demonstrate that \corpusname{} significantly facilitates LLMs in rare disease diagnosis, improving their accuracy by an average of $\ 8 \%$ on the \Algnameabbr{} dataset.


\begin{figure}[t]
    \centering
    
    \begin{minipage}[b]{1.0\linewidth}
    \centering
    \subfigure[]{
    \includegraphics[width=1.0\linewidth]{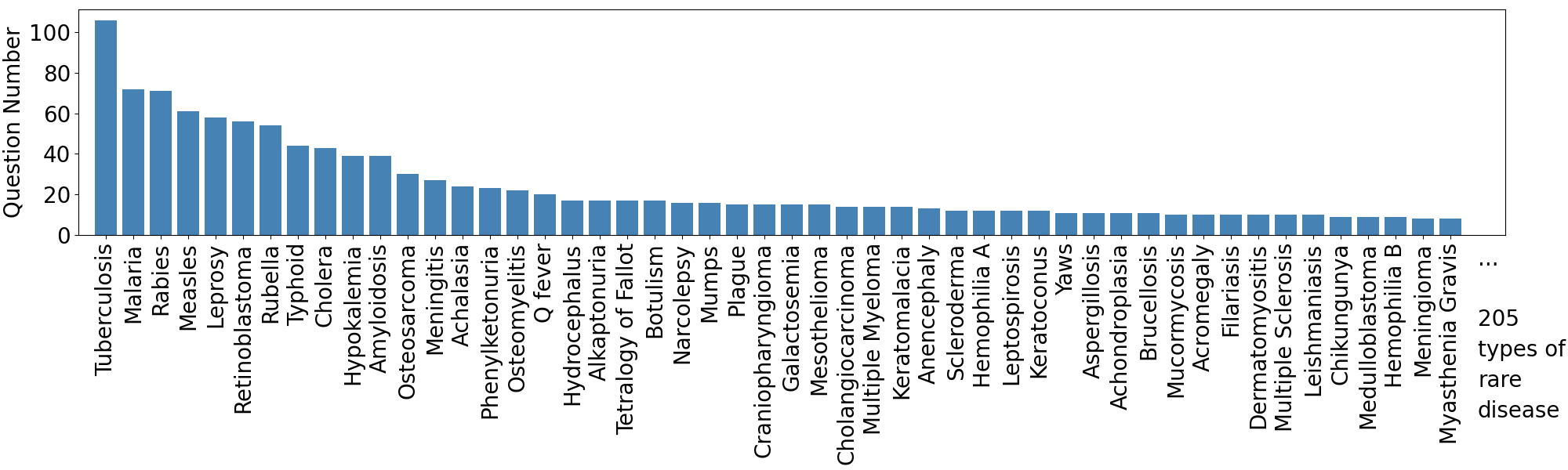}
    }
    \end{minipage}

    \begin{minipage}[b]{0.4\linewidth}
    \centering
    \subfigure[]{
        \includegraphics[width=1.0\linewidth]{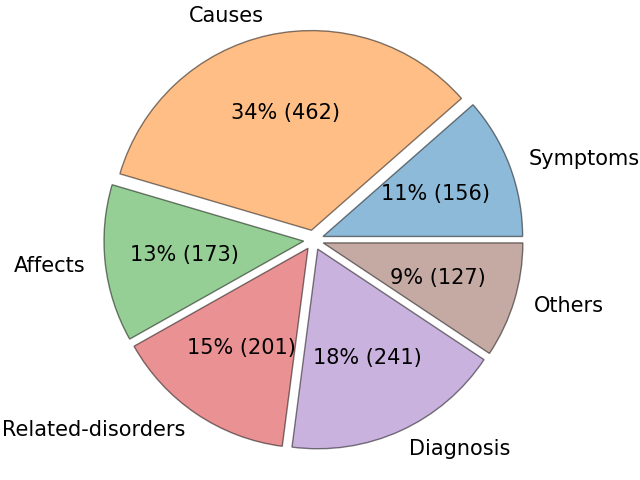}
        }
    \end{minipage}
    \begin{minipage}[b]{0.57\linewidth}
    \centering
    \subfigure[]{
        \includegraphics[width=1.0\linewidth]{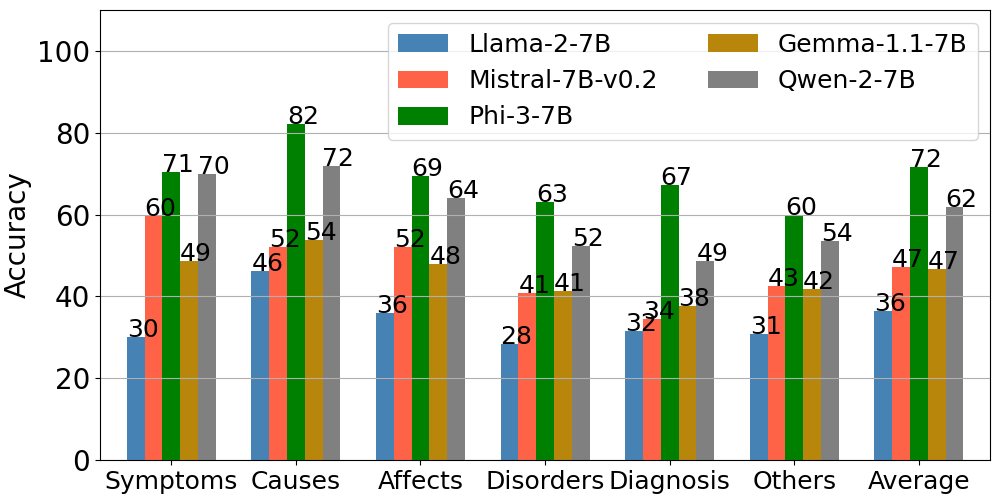}
        }
    \end{minipage}
    \caption{\label{fig:dataset_meta_data}(a) Top-50 rare diseases in the \Algnameabbr{} datasets. (b) Ratios of questions corresponding to the symptoms, causes, effects, related disorders, diagnosis, and others properties in the \Algnameabbr{} datasets. (c) Benchmark results of LLMs on the \Algnameabbr{} datasets with accuracy for each subset of properties displayed separately.}
\end{figure}

\section{Preliminaries}


\subsection{Large Language Models for Rare Disease Diagnosis.}

Large language models (LLMs) have demonstrated competitiveness in identifying rare diseases~\cite{kafkas2023application, reese2024evaluation, do2024assessing}. For instance, previous research~\cite{kafkas2023application} has shown that LLMs are effective in ranking causal genetic mutations based on their phenotypes. Another study~\cite{oniani2023large, hou2023geneturing} highlights the proficiency of LLMs in predicting patients' phenotypes from clinical notes. 
Despite the promising capabilities of LLMs in diagnosing rare diseases, most existing studies utilize closed-source frameworks and data. This presents challenges for healthcare researchers, practitioners, and enthusiasts to reproduce results or deploy these models in practice.

This motivates us to contribute to open-source efforts that benefit practitioners and researchers in practice. Specifically, we have collected and open-sourced the dataset dedicated to rare disease question-answering, \Algnameabbr{}. 
\Algnameabbr{} serves as a benchmark dataset for assessing the capabilities of LLMs in diagnosing rare diseases.
Additionally, we have open-sourced the first rare disease corpus, \corpusname{}. \corpusname{} can significantly enhance the performance of LLMs in diagnosing rare diseases through retrieval-augmented generation.



\subsection{Retrieval Augmentation Generation.}

Retrieval-Augmented Generation (RAG) can significantly enhance the performance of LLMs by leveraging retrieval-based methods to supply additional knowledge for inference.
A standard RAG framework comprises three key elements: a corpus, retriever, and LLM.
Specifically, the corpus in RAG systems serves as a comprehensive collection of knowledge. 
Retrievers play a crucial role in selecting the knowledge relevant to the question. 
For instance, dense retrievers like MedCPT~\cite{xiong2024benchmarking} match corpus chunks with questions based on the similarity of their embeddings, while sparse retrievers like BM25~\cite{robertson1995okapi} match corpus chunks with questions according to word overlap.
Finally, LLMs generate answers by using prompts based on the retrieved relevant knowledge, thereby providing more accurate and contextually appropriate answers.

RAG is particularly powerful in applications requiring access to a large body of external knowledge for generating accurate solutions. 
One typical scenario is using LLMs for diagnosing rare diseases. This task heavily depends on the knowledge of rare diseases, which may not be included in the pre-training data.
Therefore, having a corpus with high-quality knowledge is crucial for RAG systems. 
Generally, a corpus rich in rare disease-related information significantly benefits RAG in solving tasks related to rare diseases. 
Although there are existing corpora for RAG in medical tasks, such as PubMed~\cite{canese2013pubmed}, Textbook~\cite{xiong2024benchmarking}, StatPearls~\cite{StatPearls2024}, and Wikipedia~\cite{wikidump}, there is still a lack of comprehensive knowledge about rare diseases to enhance LLMs' diagnostic capabilities.
This motivates us to collect and open-source the first rare disease corpus, \corpusname{}, to improve the performance of LLMs in diagnosing rare diseases.

\section{The Rare Disease Question Answering~(\Algnameabbr{}) Dataset}

In this section, we introduce the \Algnameunderline{}~(\Algnameabbr{}) dataset in details.
We hope the \Algnameabbr{} dataset can provide a comprehensive resource for healthcare researchers, practitioners, and enthusiasts to explore the intricacies of rare diseases.
The overall framework of collecting the \Algnameabbr{} dataset is shown in Figure~\ref{fig:dataset-collection}.
The data collection pipeline includes the steps of data cleaning and labeling.


\subsubsection{Data Source and Cleaning}

The data sources for \Algnameabbr{} include the MedMCQA~\cite{pal2022medmcqa}, MedQA~\cite{jin2021disease}, and MMLU~\cite{hendrycks2020measuring} datasets. 
Specifically, MedMCQA gathers multiple-choice questions and answers from the AIIMS and NEET PG entrance exams, covering 2400 healthcare topics and 21 medical subjects. 
MedQA is a large-scale medical question-answer dataset.
The MMLU (Massive Multitask Language Understanding) dataset spans 57 subjects across STEM (Science, Technology, Engineering, and Mathematics), humanities, social sciences, and more, with difficulty levels ranging from elementary to advanced professional levels. 
In total, these sources provide over 200,000 raw question-answer pairs with high topical diversity. 
Since we target a dataset to benchmark the capabilities of LLMs in diagnosing rare diseases, the dataset cleaning process focused on removing questions irrelevant to rare diseases. 
After manual cleaning, we retained 1360 high-quality question-answer pairs relevant to rare disease diagnosis, building the \Algnameabbr{} dataset. This dataset encompasses 205 rare diseases. 
Examples of questions are shown in Figure~\ref{fig:dataset-collection}.


\begin{table}[t]
    \vspace{-2mm}
    \centering
    \footnotesize
    \begin{tabular}{l|c}
    \toprule
    & \textbf{Example} \\
    \midrule
       \!\!\!\! \rotatebox[origin=c]{90}{\textbf{Symptoms}} \!\!\!\!  &  
\makecell[l]{\textbf{Question:} About Achalasia Cardia: 
1. \textcolor{darkred}{Dysphagia is a presenting symptom}
2. The cause is the absence of Auerbach's plexus
3. Esophagec- \!\!\!\! \\
tomy is the treatment 
4. Motility improving agents are used in treatment
5. Barium swallow shows irregular filling defects in lower esophagus \!\!\!\! \\ 
\textbf{Choices:}
(A) 1,2,3 False \& 4,5 True (B) 1,2,4 True \& 3,5 False (C) 2,3,4 True \& 1,5 False (D) 1,3,5 True \& 2,4 False \\ 
\textbf{Golden Answer: B}} \\
\midrule
    \!\!\!\! \rotatebox[origin=c]{90}{\textbf{Causes}} \!\!\!\!  &  
\makecell[l]{\textbf{Question:} 
A mother brings her 1-year-old daughter to the physician. She says that for the last 2 days her daughter has been fussy and crying \!\!\!\! \\
more than usual. She also refuses formula. The patient has a fever of 39.4degC (102.9degF). Meningitis is suspected, and a lumbar puncture \!\!\!\! \\
is performed. Analysis of the cerebrospinal fluid shows an opening pressure of 98 mm H2O, a leukocyte count of 1256/mm3, a protein level \!\!\!\! \\ 
of 210 mg/dL, and a glucose level of 31 mg/dL. The mother says that the patient has received no immunizations. \textcolor{darkred}{Which of the following} \!\!\!\! \\
\textcolor{darkred}{organisms is most likely responsible for this patient's illness?} \\
\textbf{Choices:}
(A) Clostridium botulinum (B) Haemophilus influenza (C) Neisseria meningitides (D) Streptococcus pneumonia \\
\textbf{Golden Answer: B}}
\\ \midrule
    \!\!\!\! \rotatebox[origin=c]{90}{\textbf{Affects}} \!\!\!\!  &  
\makecell[l]{\textbf{Question:} 
23 years old female comes to OG, because she thinks pregnant. She missed her last two cycles and she feels different. Urine pre- \!\!\!\!\!\! \\
gnancy test was positive. On USG, the pregnancy was confirmed to be 12 weeks. She is very concerned because she received Measles Mumps \!\!\!\!\!\! \\
Rubella vaccine 4 months ago and she was told to wait for 3 months to conceive. The pregnancy is desired. The most appropriate step is \!\!\!\!\!\! \\
\textbf{Choices:}
\textcolor{darkred}{(A) Vaccine risk is minimal, not itself a reason to terminate the pregnancy. (B) Vaccine risk is nil, termination is completely} \!\!\!\!\!\! \\ 
\textcolor{darkred}{inappropriate. (C) Vaccine risk is high, termination should be strongly considered. (D) Vaccine risk is high, termination is mandated.} \!\!\!\!\!\! \\
\textbf{Golden Answer: A}}
\\ \midrule
   \!\!\!\!  \rotatebox[origin=c]{90}{\textbf{Related-disorders}} \!\!\!\!  &  
\!\! \makecell[l]{\textbf{Question:} 
\textcolor{darkred}{A 35-year-old female presented to the medicine OPD with paresthesias and weakness of B/L lower limbs with a band like sensation} \!\!\!\!\!\! \\
\!\! \textcolor{darkred}{of tightness around the torso along with painful loss of vision in both eyes along with diplopia and periorbital pain.} There is a history of similar \!\!\!\!\!\! \\
\!\! attacks in the past with period of normalcy in between O/E, Ataxia- present Papillitis (on fundus examination) Facial myokymia Bladder incon- \!\!\!\!\!\! \\ 
\!\! tinence and constipation CSF studies revealed mononuclear cell pleocytosis along with increased IgG. Which of the following are the oral drugs \!\!\!\!\!\! \\
\!\! approved for the above condition: 1. Fingolimod 2. Natalizumab 3. Teriflunomide 4. Glatiramer acetate \!\!\!\!\!\! \\
\!\! \textbf{Choices:} (A) Only 1 (B) Both 1 and 3 (C) 1,2 and 3 (D) All of the above \\
\!\! \textbf{Golden Answer: A}}
\\ \midrule
   \!\!\!\!  \rotatebox[origin=c]{90}{\textbf{Diagnosis}} \!\!\!\!  &  
\makecell[l]{\textbf{Question:} 
A 29-year-old man is seen in the office after returning from a hiking trip in Colorado. He complains of feeling unwell and reports \!\!\!\!\!\! \\
symptoms of fever, myalgia, headache, and nausea. Two days ago, he noticed a rash on his wrists and ankles that has now spread to his body. \!\!\!\!\!\! \\
He recalls having had numerous insect bites during his trip. On examination, his blood pressure is 90/60 mmHg, pulse 100/min, and respira- \!\!\!\!\!\! \\
tions 20/min. There are multiple 1-5 mm macules on his body and some of them have a hemorrhagic center consistent with a petechia. His \!\!\!\!\!\! \\
neck is supple and fundi are normal. The heart sounds are normal, lungs clear, and legs are edematous. Cranial nerve, motor, and sensory \!\!\!\!\!\! \\
examination is normal. A clinical diagnosis of Rocky Mountain Spotted Fever (RMSF) is made and he is started on appropriate therapy. \!\!\!\!\!\! \\
\textcolor{darkred}{Which of the following is the most common type of central nervous system (CNS) presentation in this condition?} \\
\textbf{Choices:} (A) Hemiplegia (B) Cranial nerve abnormalities (C) Paraplegia (D) Encephalitis
        \\
\textbf{Golden Answer: D}} \\ 
\bottomrule
    \end{tabular}
    \vspace{-2mm}
    \caption{Examples of questions and answers with different properties in the \Algnameabbr{} dataset. 
    The text in \textcolor{darkred}{dark red} provides details corresponding to the labeled property.
    }
    \label{tab:dataset_example}
\end{table}

\subsubsection{Data Labeling}

Data labeling focuses on annotating meta-data for each question-answer pair.
The meta-data includes the rare disease name and property for each question.
The rare disease name indicates the type of rare disease the question addresses. 
The property specifies the type of knowledge required to answer the question, which can be one of the following values: \emph{symptoms, causes, effects, related disorders, diagnosis}, or \emph{others}.

Based on the annotated meta-data, we show the statistics of \Algnameabbr{} dataset in Figure~\ref{fig:dataset_meta_data}.
Specifically, the top-50 rare diseases in the \Algnameabbr{} dataset are shown in Figure~\ref{fig:dataset_meta_data}~(a).
It is shown that it widely covers 205 types of rare diseases, where the most frequent disease features over 100 questions.
Regarding the property of each question, as shown in Figure~\ref{fig:dataset_meta_data}~(b), \Algnameabbr{} includes 11\%, 33\%, 13\%, 15\%, 18\% of the questions corresponding to the symptoms, causes, affects, related-disorders, diagnosis of rare diseases, respectively. 
The remaining 9\% of the questions pertain to other properties of the diseases.

To illustrate the \Algnameabbr{} dataset, here is an example question regarding the symptoms of Achalasia Cardia:
\textbf{Question:} About Achalasia Cardia: 
1. Dysphagia is a presenting symptom 
2. The cause is the absence of Auerbach's plexus 
3. Esophagectomy is the treatment 
4. Motility improving agents are used in treatment
5. Barium swallow shows irregular filling defects in lower esophagus.  \textbf{Choices:}
(A) 1,2,3 False \& 4,5 True (B) 1,2,4 True \& 3,5 False (C) 2,3,4 True \& 1,5 False (D) 1,3,5 True \& 2,4 False \textbf{Golden Answer: B}. Additional examples corresponding to the symptoms, causes, affects, related-disorders, diagnosis are shown in Table~\ref{tab:dataset_example}.



\begin{figure}[t]
    \centering
    \includegraphics[width=1.0\linewidth]{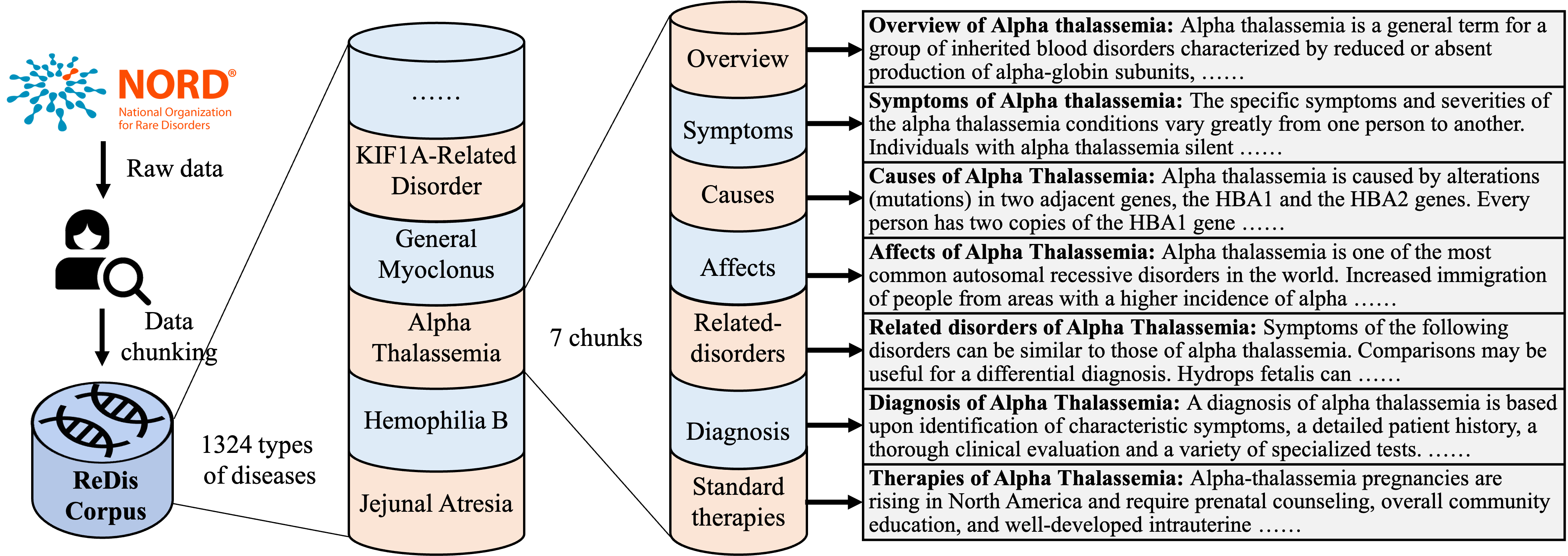}
    \caption{Pipeline of building the rare disease corpus (\corpusname{}): data collection and chunking.}
    \label{fig:corpus} 
\end{figure}

\section{Benchmark of LLMs on \Algnameabbr{} Dataset}

We benchmark open-sourced LLMs on the \Algnameabbr{} dataset to study their capabilities of rare disease diagnosis.
\paragraph{Experimental Setup.} The experiments are conducted based on the Llama-2-7B~\cite{touvron2023llama}, Mistral-7B-v0.2~\cite{jiang2024mixtral}, Phi-3-7B~\cite{abdin2024phi}, Gemma-1.1-7B~\cite{team2024gemma}, and Qwen-2-7B~\cite{bai2023qwen} LLMs.
These LLMs represent the state-of-the-art in natural language question answering.
We load the instruction tuned versions of these models from the Huggingface platform~\cite{wolf2019huggingface}.
The evaluation metric is the accuracy on the \Algnameabbr{} dataset.
The prompts for LLMs are given in Appendix~\ref{alg:eb_mc_rag}.

\paragraph{Benchmark Results.} Figure~\ref{fig:dataset_meta_data}~(b) illustrates the accuracy of LLMs in percentage, with accuracy for each subset of properties displayed separately. We have the following observations:
\begin{itemize}

    \item \textbf{Overall Performance.} Phi-3-7B shows the most competitive performance across all properties, while Llama-2-7B, Mistral-7B-v0.2, and Gemma-1.1-7B are less qualified than Phi-3-7B and Qwen-2-7B in rare disease diagnosis because they have accuracy less than 50\% for all properites.

    \item \textbf{Easy and Hard Properties.} LLMs show higher accuracy on questions of rare diseases' symptoms, causes, and effects, while exhibiting less qualification on questions of related disorders, diagnosis, and other properties.
    

\end{itemize}

\section{The Rare Disease Corpus}

The data for \corpusname{} is sourced from the National Organization for Rare Disorders (NORD) database\footnote{\url{https://rarediseases.org/}}, which compiles reports on rare diseases. 
\emph{NORD is committed to the identification, treatment, and cure of rare diseases through education, advocacy, research, and service programs.}
The primary objective of developing \corpusname{} using the NORD database is to provide comprehensive expertise on rare diseases for LLMs. 
This expertise can be leveraged to enhance the diagnostic capabilities of LLMs through retrieval-augmented generation.
The pipeline for constructing \corpusname{} includes: data collection and data chunking steps, as illustrated in Figure~\ref{fig:corpus}.





\begin{table}[t]
    \centering
    \caption{Accuracy (\%) of LLMs with and without \corpusname{} on the \Algnameabbr{} dataset.}
    \label{tab:rag_results_main}
    \resizebox{\textwidth}{!}{
    \begin{tabular}{l|c|ccccccc}
    \toprule
         Retriever & $k$ & Llama-2-7B & Mistral-7B-v0.2 & Phi-3-7B & Gemmma-1.1-7B & Qwen-2-7B & Average & Improve \\
    \midrule
        N/A & 0 & 36.4 & 47.3 & 71.6 & 46.6 & 61.9 & 52.8 & 0 \\
    \midrule
        \multirow{2}{*}{Meta-data} & 5 & 40.8 & 57.1 & 74.6 & 58.2 & 66.2 & 59.4 & +6.6 \\
        & 7 & 39.8 & \textbf{60.9} & \textbf{75.1} & \textbf{61.2} & \textbf{67.6} & \textbf{60.9} & \textbf{+8.1} \\
    \midrule
        \multirow{2}{*}{MedCPT} & 5 & 42.9 & {55.0} & {72.4} & {57.2} & 65.1 & 58.5 & +5.7 \\
        & 7 & \textbf{43.1} & 54.4 & 71.8 & 55.8 & {65.9} & 58.2 & +5.4 \\
    \midrule
        \multirow{2}{*}{BM25} & 5 & {41.9} & 55.3 & 72.1 & {53.8} & {64.9} & 57.6 & +4.8 \\
        & 7 & 41.5 & {55.5} & {72.5} & 53.5 & 63.8 & 57.4 & +4.6 \\
    \bottomrule
    \end{tabular}
    }
\end{table}

\subsubsection{Data Source}

The data for \corpusname{} is sourced from the National Organization for Rare Disorder database, which contains professional reports on 1324 rare diseases. 
Each report includes comprehensive information on the symptoms, causes, effects, treatments, and clinical trials related to a specific rare disease. 
The reports are written in non-technical language, making them accessible to both non-professional individuals and LLMs. 
Additionally, NORD provides several details that enhance the ability of LLMs in rare disease diagnosis:

\begin{itemize}

    \item \textbf{Genetic mutations.} Each report details the specific genetic mutation causing the disease, if applicable. For example, Klinefelter Syndrome is caused by an extra X chromosome in cells.

    \item \textbf{Synonyms of diseases.} Each report lists the synonyms of the disease to prevent misunderstandings that may arise from varying terminologies used in different literature.

    \item \textbf{Reference.} NORD includes references to scientific articles, textbooks, and government agency reports for each report, ensuring the data's reliability and trustworthiness.
    
\end{itemize}



\subsubsection{Data Chunking}

Data chunking is a crucial step for retrieval augmentation. Chunks are the minimal units used to match queries and provide prompts during retrieval augmentation. 
A good chunking strategy ensures consistent relations between documents and queries, where a document is either relevant or irrelevant to the query, thus avoiding the issue where parts of a document are relevant while other parts are not.
We consider the metadata of the \Algnameabbr{} dataset to chunk the reports of each rare disease in the NORD database. To align with this metadata, \corpusname{} divides each rare disease report into chunks: \emph{overview, symptoms, causes, effects, related disorders, diagnosis}, and \emph{standard therapies}. Each property of the disease corresponds to a specific chunk in \corpusname{}.
For example, Figure~\ref{fig:corpus} illustrates the chunks for the report on alpha thalassemia. In this manner, \corpusname{} generates 9268 chunks from the NORD database, with each report producing seven chunks corresponding to the properties of a rare disease.


\section{Benchmark of Retrieval Augmentation with \corpusname{} on \Algnameabbr{} Dataset}


We demonstrate that \corpusname{} can significantly improve LLM performance in rare disease QA by contributing to the retrieval augmentation generation (RAG) of LLMs. 
The experiments are based on the \Algnameabbr{} dataset.


\paragraph{Experimental Setup.}
The prerequisites for the experiments include LLMs, retrieval algorithms, and corpus data. 
Specifically, we use the Llama-2-7B, Mistral-7B-v0.2, Phi-3-7B, Gemma-1.1-7B, and Qwen-2-7B LLMs, consistent with the previous LLM benchmark experiments. For retrieval algorithms, we consider the meta-data retriever, MedCPT~(dense retriever)~\cite{xiong2024benchmarking}, and BM25~(sparse retriever)~\cite{robertson1995okapi}. 
Concretely, the meta-data retriever matches questions and \corpusname{} chunks using rare disease names as keywords; MedCPT matches the questions with corpus chunks by their embeddings; and BM25 matches them by overlapped words.
Additionally, we use the PubMed~\cite{canese2013pubmed}, Textbook~\cite{xiong2024benchmarking}, StatPearls~\cite{StatPearls2024}, and Wikipedia~\cite{wikidump} databases as baseline corpus for comparison.
The prompts for LLMs w/o and w/ RAG are given in Appendix~\ref{appendix:prompts}.
The algorithm of combing baseline corpus with \corpusname{} is given in Appendix~\ref{appendix:comb_alg}.

\begin{figure}
\centering
\subfigure[Mistral-7B]{
\centering
	\begin{minipage}[t]{0.25\linewidth}
		\includegraphics[width=0.99\linewidth]{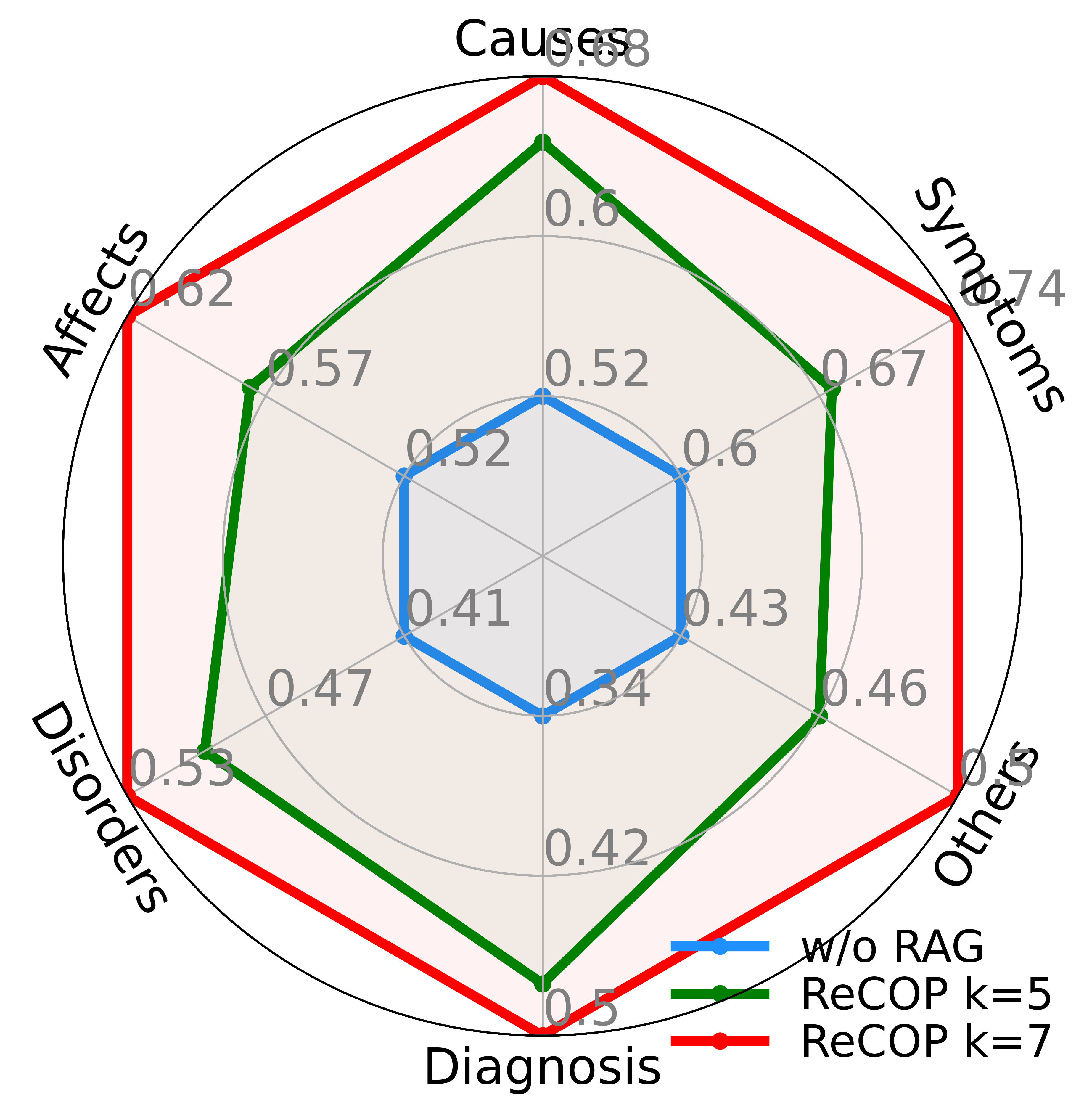}
	\end{minipage}
}
\!\!\!\!\!\!
\subfigure[Phi-3-7B]{
\centering
	\begin{minipage}[t]{0.25\linewidth}
		\includegraphics[width=0.99\linewidth]{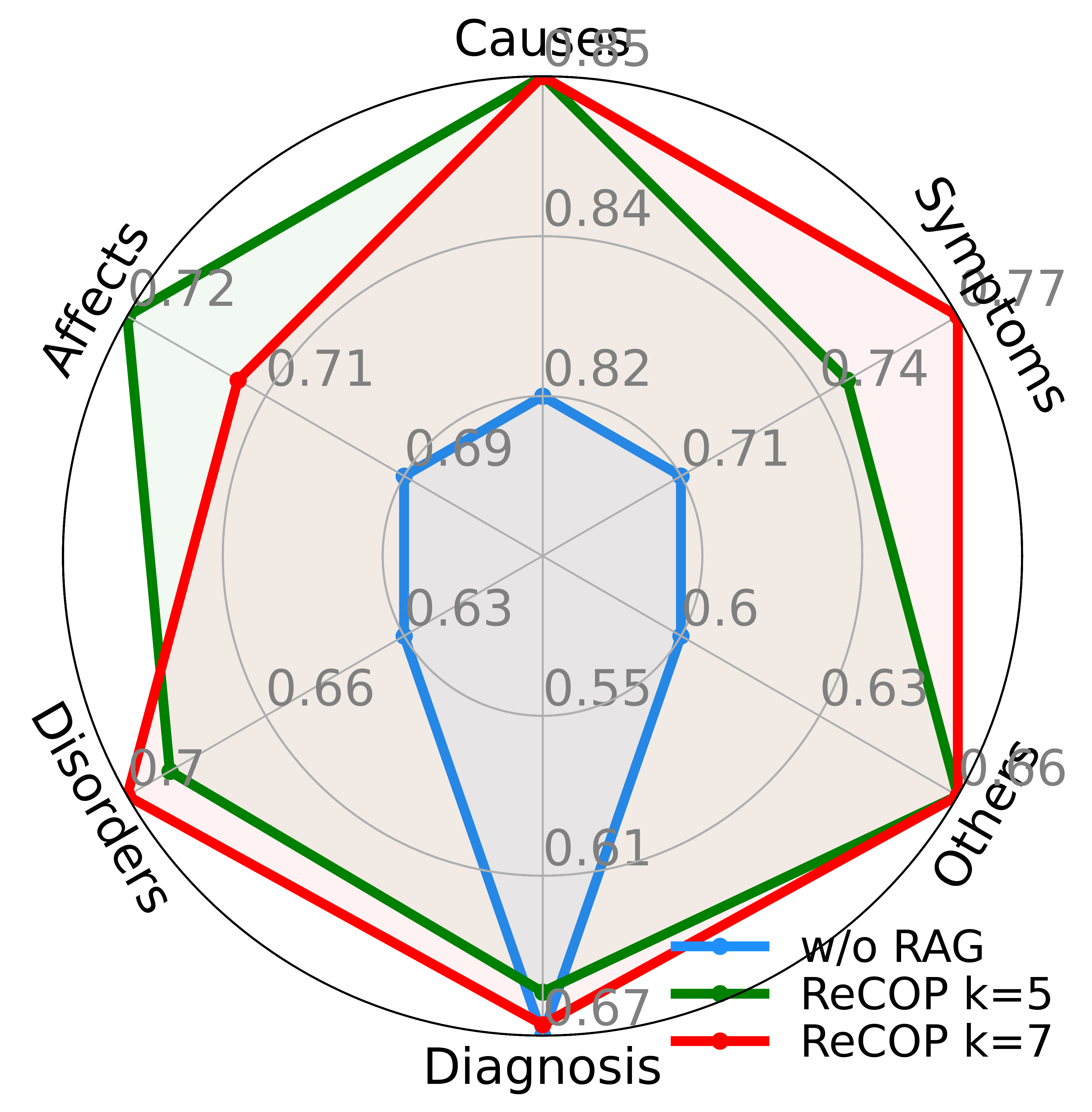}
	\end{minipage}%
}
\!\!\!\!\!\!
\subfigure[Gemma-1.1-7B]{
\centering
	\begin{minipage}[t]{0.25\linewidth}
		\includegraphics[width=0.99\linewidth]{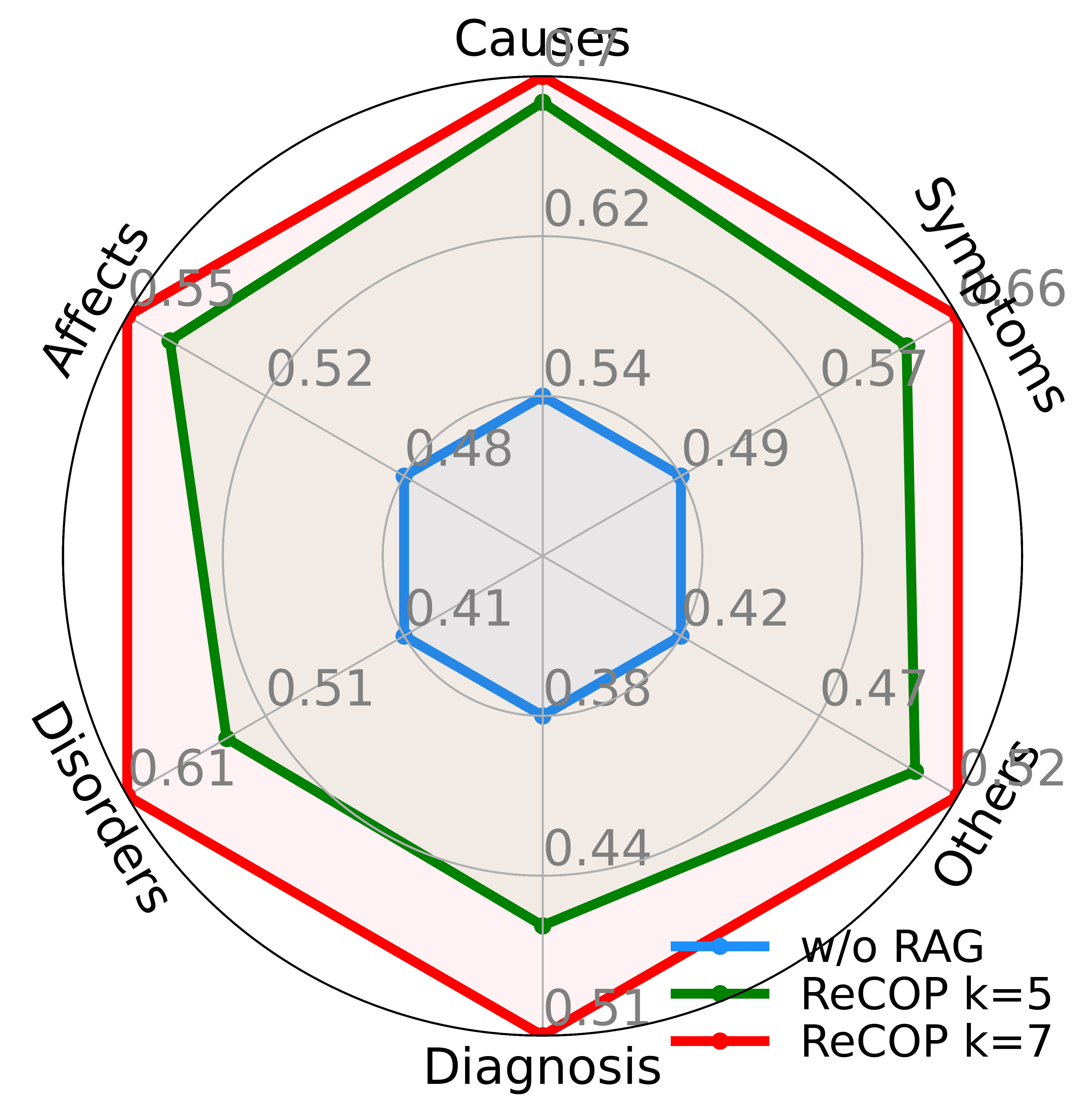}
	\end{minipage}%
}
\!\!\!\!\!\!
\subfigure[Qwen-2-7B]{
\centering
	\begin{minipage}[t]{0.25\linewidth}
		\includegraphics[width=0.99\linewidth]{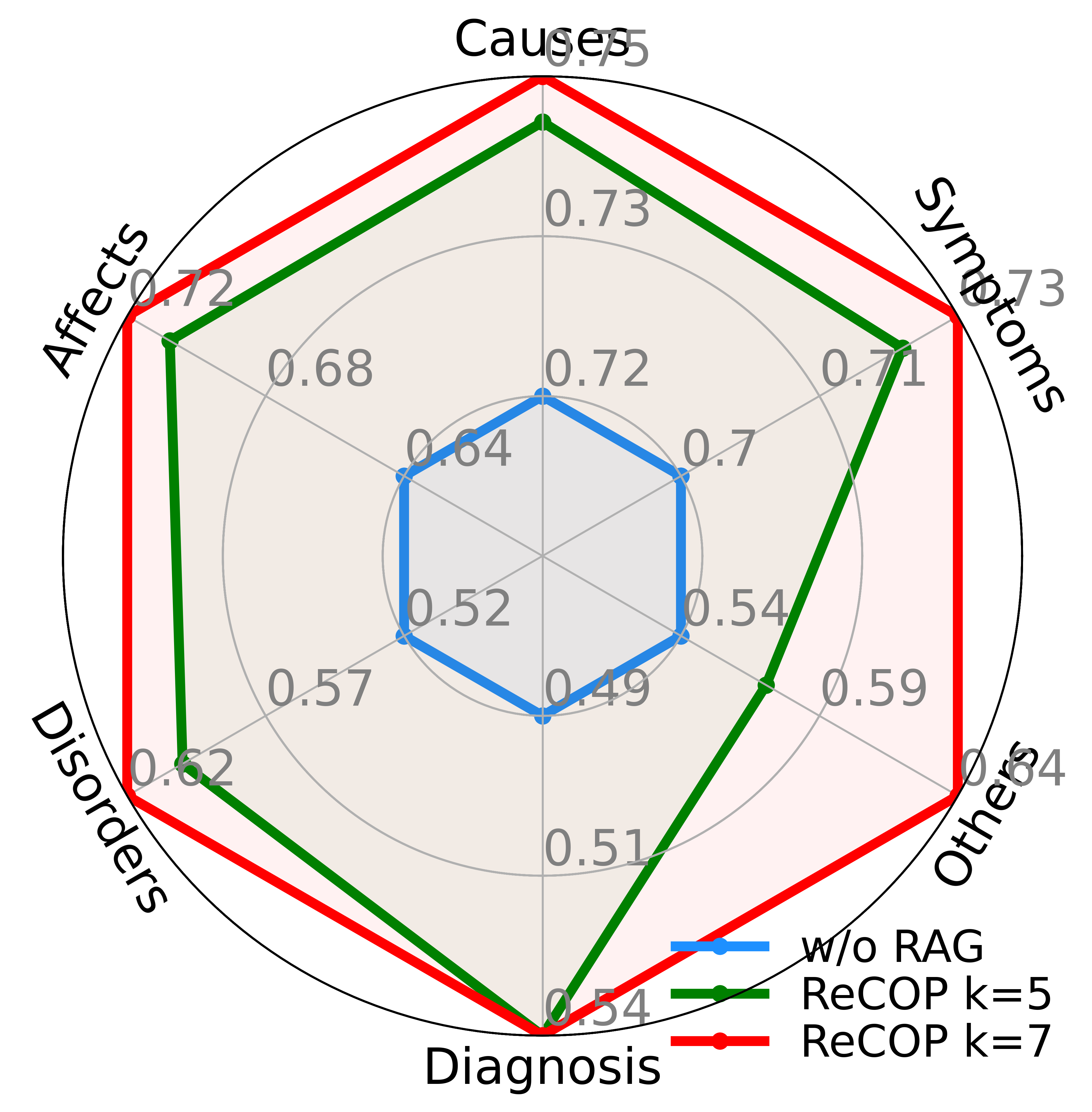}
	\end{minipage}%
}
\centering
\subfigure[\corpusname{}+Textbook]{
\centering
	\begin{minipage}[t]{0.25\linewidth}
		\includegraphics[width=0.99\linewidth]{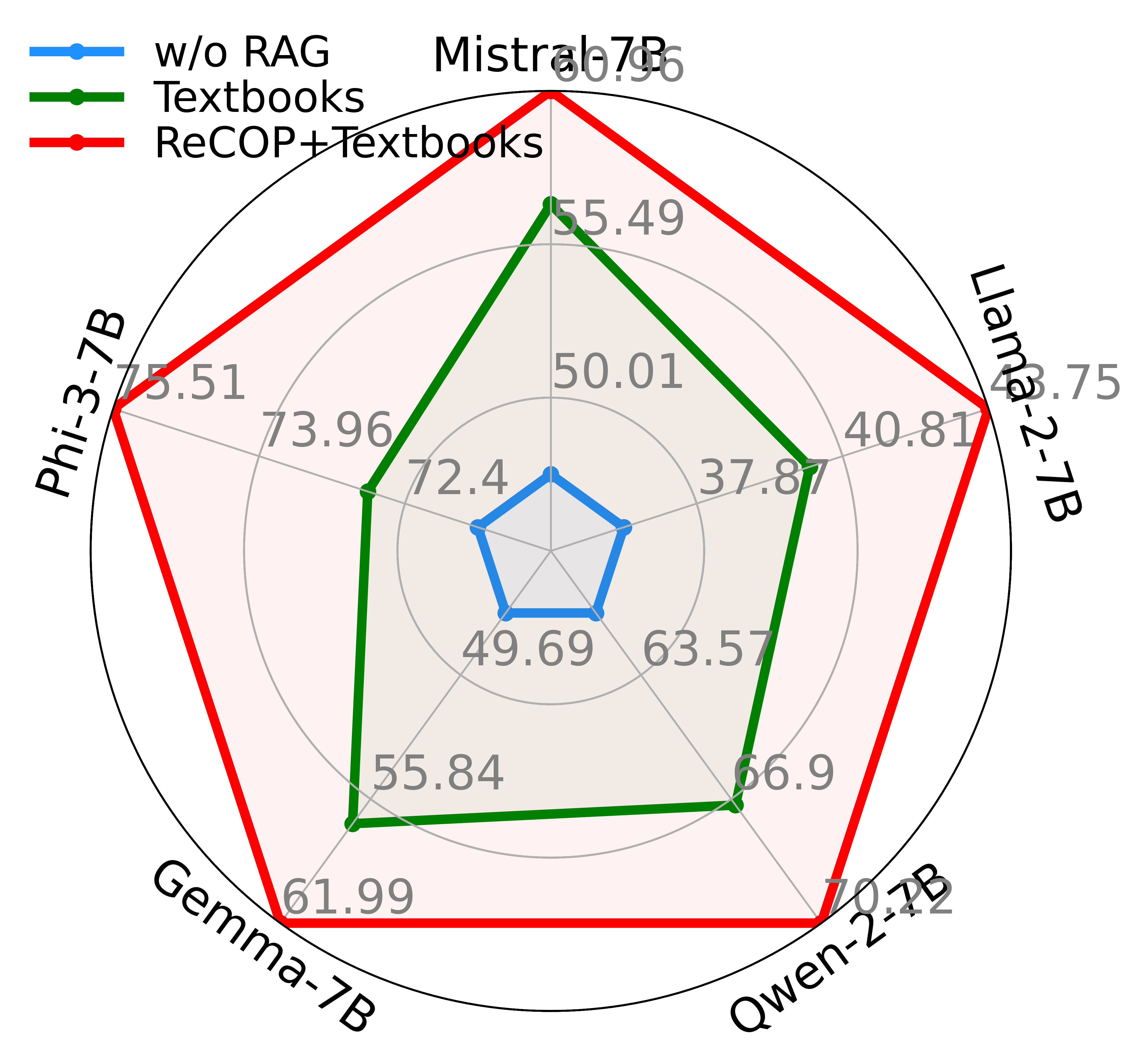}
	\end{minipage}
}
\!\!\!\!\!\!
\subfigure[\corpusname{}+StatPearls]{
\centering
	\begin{minipage}[t]{0.25\linewidth}
		\includegraphics[width=0.99\linewidth]{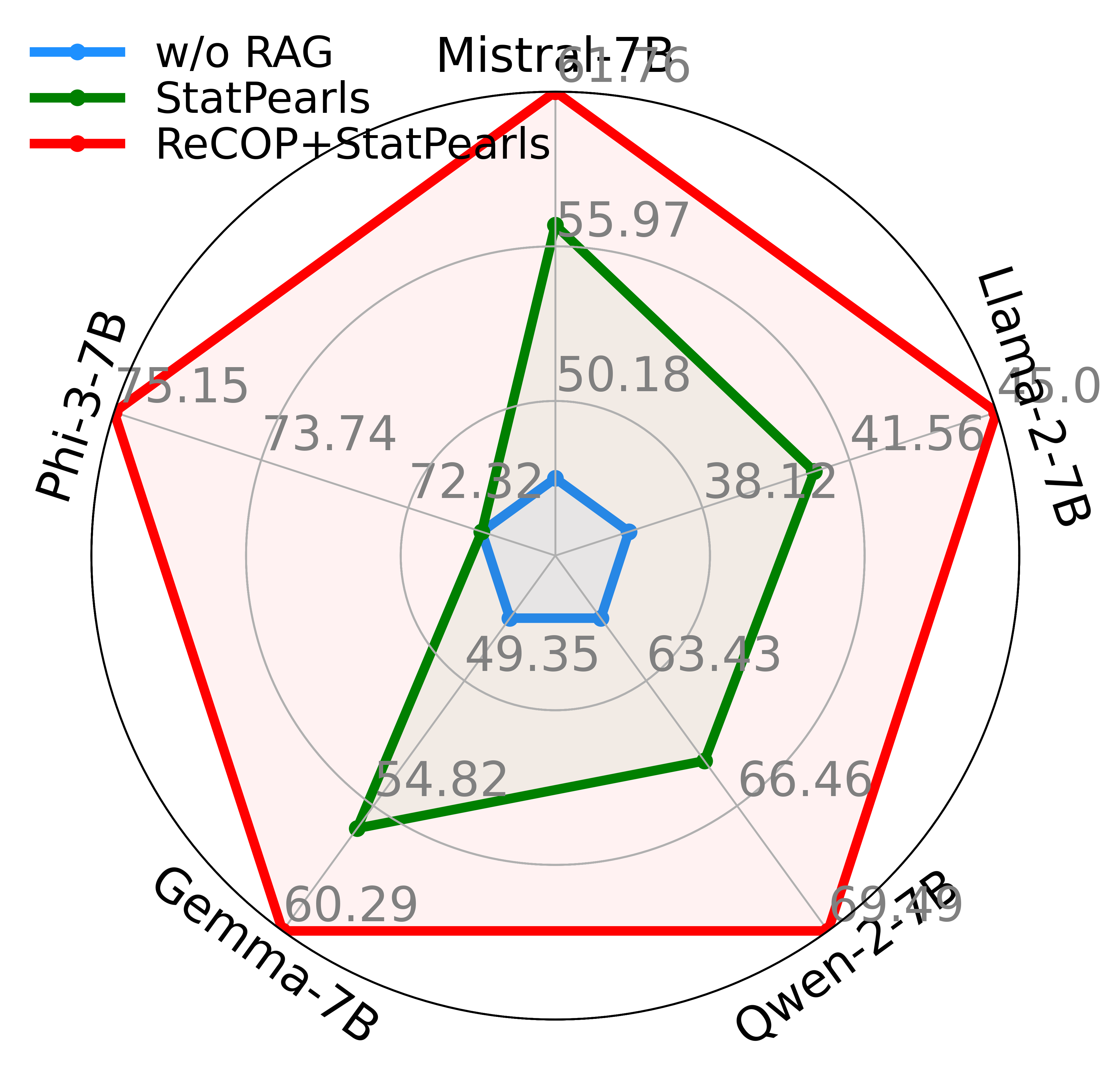}
	\end{minipage}%
}
\!\!\!\!\!\!
\subfigure[\corpusname{}+PubMed]{
\centering
	\begin{minipage}[t]{0.25\linewidth}
		\includegraphics[width=0.99\linewidth]{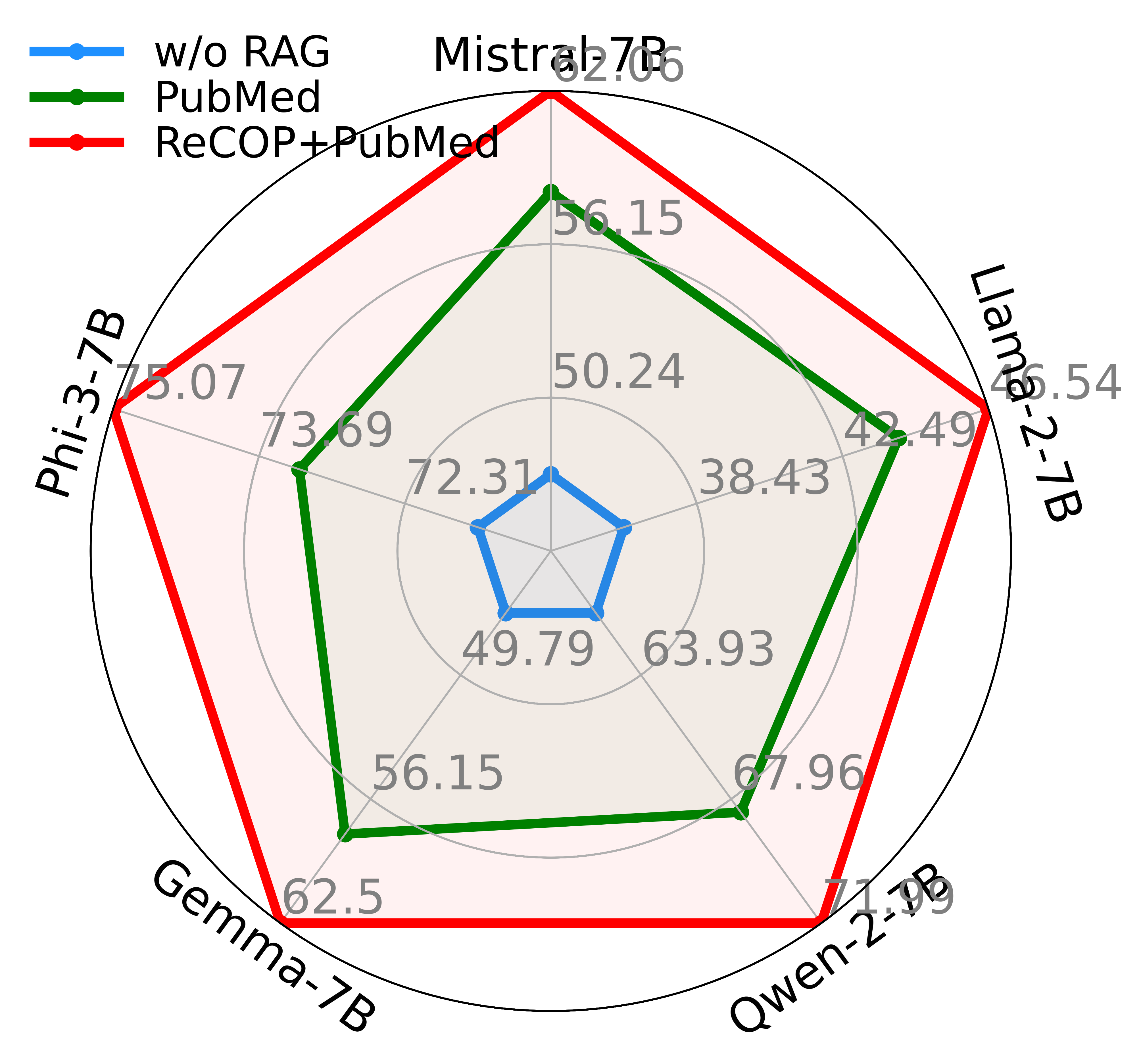}
	\end{minipage}%
}
\!\!\!\!\!\!
\subfigure[\corpusname{}+Wikipedia]{
\centering
	\begin{minipage}[t]{0.25\linewidth}
		\includegraphics[width=0.99\linewidth]{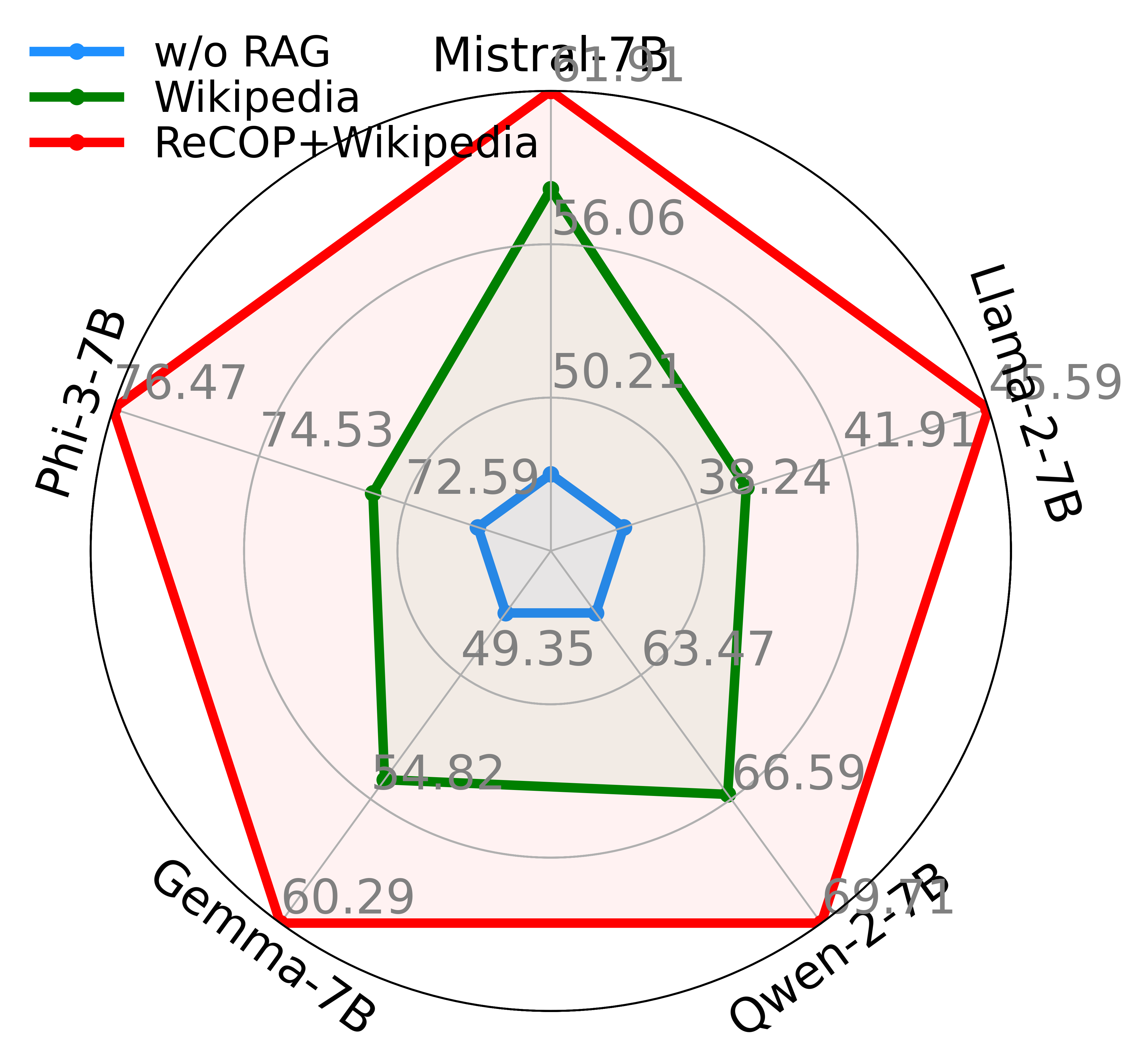}
	\end{minipage}%
}
\caption{\small (a)-(d) Accuracy of LLMs with and without \corpusname{} on the six subsets of disease properties.
(e)-(h) Accuracy of LLMs without \corpusname{}, LLMs with RAG using baseline corpus, and LLMs with RAG using baseline corpus and \corpusname{}, where baseline corpous take Textbooks~(e), StatPearls~(f), PubMed~(g), and Wikipedia~(h); the retriever takes BM25 with $k=7$.}
\label{fig:radar_plot}
\end{figure}


\paragraph{Benchmark Results.}

Table \ref{tab:rag_results_main} shows the accuracy of LLMs with and without RAG, where $k$ indicates the number of retrieved chunks for prompting LLM inferences.
Overall, we have the following observations:
\begin{itemize}

    \item \textbf{LLMs w/ \corpusname{} vs. LLMs w/o RAG} Table \ref{tab:rag_results_main} demonstrates that LLM w/ \corpusname{} outperforms LLMs w/o RAG by an average of $8\%$. 
    Figure~\ref{fig:radar_plot}~(a)-(d) further illustrates that LLMs w/ \corpusname{} exceeds LLMs w/o RAG performance across all six properties of the questions. 
    These comprehensive results highlight the effectiveness of \corpusname{} in providing external knowledge to enhance LLM performance in rare disease question answering.

    \item \textbf{Meta-data Retriever vs. MedCPT and BM25.} Table \ref{tab:rag_results_main} shows that the meta-data retriever outperforms MedCPT and BM25. This indicates the labeled meta-data of \Algnameabbr{} and chunking of \corpusname{} provide useful prior knowledge for LLMs in rare disease question answering.

    \item \textbf{\corpusname{} vs. Textbook and StatPearls.} \corpusname{} offers unique knowledge distinct from the baseline corpus for LLMs. 
    As illustrated in Figure~\ref{fig:radar_plot}~(e) and (f), \corpusname{} combined with either the Textbook, StatPearls, PubMed, and Wikipedia corpus significantly enhances LLMs' performance in diagnosing rare diseases. 
    Integrating \corpusname{} with other corpus significantly enriches the knowledge for LLMs improving their diagnostic capabilities for rare diseases. More results of combining \corpusname{} with baseline corpus are provided in Appendix~\ref{appendix:pubmed_results}.

\end{itemize}

\section{Case Studies on Natural Language Explanation}

Natural language explanations are crucial for medical-related tasks.
As noted in the literature~\cite{pavlus2019new}, ``If a doctor told that you needed surgery, you would want to know why."
Following this motivation, we investigate the explanations provided by LLMs for their answers. 
A case result is shown in Figure~\ref{fig:explanation_case}.
It is observed that LLMs without RAG follow an incorrect inference process and produce wrong answers. 
In contrast, \corpusname{} not only guides LLMs to the correct answer but also ensures they follow the correct reasoning process, with explanations that can be traced back to existing literature on Microsomal Triglyceride Transfer Protein~\cite{hussain2012multiple}. Additional cases of natural language explanations are provided in Appendix~\ref{appendix:explanation}.
By providing informative and rare disease-related knowledge for LLMs, \corpusname{} significantly enhances their trustworthiness, thereby promoting their application in real-world diagnostic scenarios.

\begin{figure*}
\fcolorbox{black}{lightgray!15}{\parbox{.99\linewidth}
{
\textbf{Question:} 
Abetalipoproteinemia is due to deficiency of 
\\
\textbf{Choices:}
(A) Lecithin Cholesterol Acyl Transferase (B) ATP Binding Cassette Transporter-1 (C) Mitochondrial Triglyceride Transfer Protein (D) ApoCII 
\vspace{2mm}
\\
\textbf{LLMs w/o RAG Answer:} \!\!\!\! \colorbox{lightred}{A. Lecithin Cholesterol Acyl Transferase (LCAT).} \textbf{Explanation:} Abetalipoproteinemia is a rare genetic disorder characterized by the absence or near absence of apoB-containing lipoproteins, including low-density lipoproteins (LDL) and very low-density lipoproteins (VLDL), in the blood. \colorbox{lightred}{This condition results from a} \colorbox{lightred}{deficiency of lecithin-cholesterol acyltransferase (LCAT),} an enzyme that plays a crucial role in the formation and maturation of these lipoproteins.
\vspace{2mm}
\\
\textbf{\corpusname{}-RAG Answer:} \!\!\!\! \colorbox{lightgreen}{C. Microsomal Triglyceride Transfer Protein (MTP).} \textbf{Explanation:} \!\!\!\! \colorbox{lightgreen}{Abetalipoproteinemia} \colorbox{lightgreen}{is caused by deficiency of microsomal triglyceride transfer protein (MTP), which is required for the proper} \colorbox{lightgreen}{assembly and secretion of apoB-containing lipoproteins in the liver and intestines.}
\vspace{2mm}
\\
\textbf{Golden Answer:} \!\!\!\! \colorbox{lightgreen}{C. Microsomal Triglyceride Transfer Protein (MTP).} \textbf{Explanation from Literature:} \!\!\!\! \colorbox{lightgreen}{MTP} \colorbox{lightgreen}{deficiency results in abetalipoproteinemia characterized by the absence of apoB-containing lipoproteins~\cite{hussain2012multiple}.}
}}
\vspace{-2mm}
\caption{\label{fig:explanation_case} Explanations provided by LLMs for the answers.}
\end{figure*}




\section{Conclusion}

In this work, we collect and open-source a rare disease question-answering dataset, \Algnameabbr{}, for benchmark the capabilities of LLMs in this area.
Specifically, it includes 1360 high-quality question-answer pairs covering 205 rare diseases. 
Each question is annotated with meta-data, facilitating the extraction of
subsets specific to any given disease and its property. 
Based on the \Algnameabbr{} dataset, we benchmark the performance existing open-source LLMs, revealing that the diagnosis of rare diseases remains a significant challenge for them.
To improve their performance, we collect and open-source a rare disease corpus, \corpusname{}, to facilitate retrieval augmentation generation for rare disease diagnosis.
Specifically, each data chunk in the \corpusname{} represents a different property of a disease, including their overview, symptoms, causes, effects, related disorders, diagnosis, and standard therapies.
Experimental results demonstrate that \corpusname{} provides unique knowledge of rare diseases that is distinct from the existing corpus, significantly improving the accuracy of LLMs on the \corpusname{} dataset.
Moreover, it significantly guilds LLMs to generate trustworthy answers and explanations that can be traced back to existing literature.



\bibliographystyle{vancouver}
\bibliography{amia} 

\begin{thebibliography}{10}

\bibitem{clusmann2023future}
Clusmann J, Kolbinger FR, Muti HS, Carrero ZI, Eckardt JN, Laleh NG, et~al.
\newblock The future landscape of large language models in medicine.
\newblock Communications medicine. 2023;3(1):141.

\bibitem{thirunavukarasu2023large}
Thirunavukarasu AJ, Ting DSJ, Elangovan K, Gutierrez L, Tan TF, Ting DSW.
\newblock Large language models in medicine.
\newblock Nature medicine. 2023;29(8):1930-40.

\bibitem{yuan2023large}
Yuan J, Tang R, Jiang X, Hu X.
\newblock Large language models for healthcare data augmentation: An example on patient-trial matching.
\newblock In: AMIA Annual Symposium Proceedings. vol. 2023. American Medical Informatics Association; 2023. p. 1324.

\bibitem{yang2023large}
Yang R, Tan TF, Lu W, Thirunavukarasu AJ, Ting DSW, Liu N.
\newblock Large language models in health care: Development, applications, and challenges.
\newblock Health Care Science. 2023;2(4):255-63.

\bibitem{bhayana2024chatbots}
Bhayana R.
\newblock Chatbots and large language models in radiology: a practical primer for clinical and research applications.
\newblock Radiology. 2024;310(1):e232756.

\bibitem{chuang2024spec}
Chuang YN, Tang R, Jiang X, Hu X.
\newblock SPeC: a soft prompt-based calibration on performance variability of large language model in clinical notes summarization.
\newblock Journal of Biomedical Informatics. 2024;151:104606.

\bibitem{touvron2023llama}
Touvron H, Lavril T, Izacard G, Martinet X, Lachaux MA, Lacroix T, et~al.
\newblock LLaMA: Open and Efficient Foundation Language Models.
\newblock arXiv preprint arXiv:230213971. 2023.

\bibitem{jiang2024mixtral}
Jiang AQ, Sablayrolles A, Roux A, Mensch A, Savary B, Bamford C, et~al.
\newblock Mixtral of experts.
\newblock arXiv preprint arXiv:240104088. 2024.

\bibitem{abdin2024phi}
Abdin M, Jacobs SA, Awan AA, Aneja J, Awadallah A, Awadalla H, et~al.
\newblock Phi-3 technical report: A highly capable language model locally on your phone.
\newblock arXiv preprint arXiv:240414219. 2024.

\bibitem{team2024gemma}
Team G, Mesnard T, Hardin C, Dadashi R, Bhupatiraju S, Pathak S, et~al.
\newblock Gemma: Open models based on gemini research and technology.
\newblock arXiv preprint arXiv:240308295. 2024.

\bibitem{wu2024pmc}
Wu C, Lin W, Zhang X, Zhang Y, Xie W, Wang Y.
\newblock PMC-LLaMA: toward building open-source language models for medicine.
\newblock Journal of the American Medical Informatics Association. 2024:ocae045.

\bibitem{xie2024me}
Xie Q, Chen Q, Chen A, Peng C, Hu Y, Lin F, et~al.
\newblock Me llama: Foundation large language models for medical applications.
\newblock arXiv preprint arXiv:240212749. 2024.

\bibitem{labrak2024biomistral}
Labrak Y, Bazoge A, Morin E, Gourraud PA, Rouvier M, Dufour R.
\newblock Biomistral: A collection of open-source pretrained large language models for medical domains.
\newblock arXiv preprint arXiv:240210373. 2024.

\bibitem{nguengang2020estimating}
Nguengang~Wakap S, Lambert DM, Olry A, Rodwell C, Gueydan C, Lanneau V, et~al.
\newblock Estimating cumulative point prevalence of rare diseases: analysis of the Orphanet database.
\newblock European journal of human genetics. 2020;28(2):165-73.

\bibitem{mao2024ai}
Mao D, Liu C, Wang L, AI-Ouran R, Deisseroth C, Pasupuleti S, et~al.
\newblock AI-MARRVEL—A Knowledge-Driven AI System for Diagnosing Mendelian Disorders.
\newblock NEJM AI. 2024;1(5):AIoa2300009.

\bibitem{stark2023genomic}
Stark Z, Scott RH.
\newblock Genomic newborn screening for rare diseases.
\newblock Nature Reviews Genetics. 2023;24(11):755-66.

\bibitem{kafkas2023application}
Kafkas {\c{S}}, Abdelhakim M, Althagafi A, Toonsi S, Alghamdi M, Schofield PN, et~al.
\newblock The application of Large Language Models to the phenotype-based prioritization of causative genes in rare disease patients.
\newblock medRxiv. 2023:2023-11.

\bibitem{yu2023leveraging}
Yu P, Xu H, Hu X, Deng C.
\newblock Leveraging generative AI and large Language models: a Comprehensive Roadmap for Healthcare Integration.
\newblock In: Healthcare. vol.~11. MDPI; 2023. p. 2776.

\bibitem{brown2020language}
Brown T, Mann B, Ryder N, Subbiah M, Kaplan JD, Dhariwal P, et~al.
\newblock Language models are few-shot learners.
\newblock Advances in neural information processing systems. 2020;33:1877-901.

\bibitem{jahan2024comprehensive}
Jahan I, Laskar MTR, Peng C, Huang JX.
\newblock A comprehensive evaluation of large language models on benchmark biomedical text processing tasks.
\newblock Computers in biology and medicine. 2024;171:108189.

\bibitem{nori2023capabilities}
Nori H, King N, McKinney SM, Carignan D, Horvitz E.
\newblock Capabilities of gpt-4 on medical challenge problems.
\newblock arXiv preprint arXiv:230313375. 2023.

\bibitem{nori2023can}
Nori H, Lee YT, Zhang S, Carignan D, Edgar R, Fusi N, et~al.
\newblock Can generalist foundation models outcompete special-purpose tuning? case study in medicine.
\newblock arXiv preprint arXiv:231116452. 2023.

\bibitem{reese2024evaluation}
Reese JT, Chimirri L, Danis D, Caufield JH, Wissink KW, Casiraghi E, et~al.
\newblock Evaluation of the Diagnostic Accuracy of GPT-4 in Five Thousand Rare Disease Cases.
\newblock medRxiv. 2024:2024-07.

\bibitem{do2024assessing}
do~Olmo J, Logrono J, Mascias C, Martinez M, Isla J.
\newblock Assessing DxGPT: Diagnosing Rare Diseases with Various Large Language Models.
\newblock medRxiv. 2024:2024-05.

\bibitem{oniani2023large}
Oniani D, Hilsman J, Dong H, Gao F, Verma S, Wang Y.
\newblock Large language models vote: Prompting for rare disease identification.
\newblock arXiv preprint arXiv:230812890. 2023.

\bibitem{hou2023geneturing}
Hou W, Ji Z.
\newblock GeneTuring tests GPT models in genomics.
\newblock BioRxiv. 2023.

\bibitem{xiong2024benchmarking}
Xiong G, Jin Q, Lu Z, Zhang A.
\newblock Benchmarking retrieval-augmented generation for medicine.
\newblock arXiv preprint arXiv:240213178. 2024.

\bibitem{robertson1995okapi}
Robertson SE, Walker S, Jones S, Hancock-Beaulieu MM, Gatford M, et~al.
\newblock Okapi at TREC-3.
\newblock Nist Special Publication Sp. 1995;109:109.

\bibitem{canese2013pubmed}
Canese K, Weis S.
\newblock PubMed: the bibliographic database.
\newblock The NCBI handbook. 2013;2(1).

\bibitem{StatPearls2024}
StatPearls. StatPearls;.
\newblock Available from: \url{https://www.statpearls.com/}.

\bibitem{wikidump}
Foundation W. Wikimedia Downloads;.
\newblock Available from: \url{https://dumps.wikimedia.org}.

\bibitem{pal2022medmcqa}
Pal A, Umapathi LK, Sankarasubbu M.
\newblock Medmcqa: A large-scale multi-subject multi-choice dataset for medical domain question answering.
\newblock In: Conference on health, inference, and learning. PMLR; 2022. p. 248-60.

\bibitem{jin2021disease}
Jin D, Pan E, Oufattole N, Weng WH, Fang H, Szolovits P.
\newblock What disease does this patient have? a large-scale open domain question answering dataset from medical exams.
\newblock Applied Sciences. 2021;11(14):6421.

\bibitem{hendrycks2020measuring}
Hendrycks D, Burns C, Basart S, Zou A, Mazeika M, Song D, et~al.
\newblock Measuring massive multitask language understanding.
\newblock arXiv preprint arXiv:200903300. 2020.

\bibitem{bai2023qwen}
Bai J, Bai S, Chu Y, Cui Z, Dang K, Deng X, et~al.
\newblock Qwen technical report.
\newblock arXiv preprint arXiv:230916609. 2023.

\bibitem{wolf2019huggingface}
Wolf T, Debut L, Sanh V, Chaumond J, Delangue C, Moi A, et~al.
\newblock Huggingface's transformers: State-of-the-art natural language processing.
\newblock arXiv preprint arXiv:191003771. 2019.

\bibitem{pavlus2019new}
Pavlus J.
\newblock A new approach to understanding how machines think.
\newblock Quanta Magazine. 2019;10.

\bibitem{hussain2012multiple}
Hussain MM, Rava P, Walsh M, Rana M, Iqbal J.
\newblock Multiple functions of microsomal triglyceride transfer protein.
\newblock Nutrition \& metabolism. 2012;9:1-16.

\bibitem{carrasco2016cutaneous}
Carrasco-Zuber J, Navarrete-Dechent C, Bonifaz A, Fich F, Vial-Letelier V, Berroeta-Mauriziano D.
\newblock Cutaneous involvement in the deep mycoses: a review. Part II—Systemic mycoses.
\newblock Actas Dermo-Sifiliogr{\'a}ficas (English Edition). 2016;107(10):816-22.

\bibitem{ornitz2017achondroplasia}
Ornitz DM, Legeai-Mallet L.
\newblock Achondroplasia: Development, pathogenesis, and therapy.
\newblock Developmental dynamics. 2017;246(4):291-309.

\bibitem{hongo1982clinical}
Hongo M.
\newblock Clinical effect of nifedipine in patients with achalasia.
\newblock Nihon Heikatsukin Gakkai Zasshi. 1982;18(1):39-43.

\bibitem{seo2018wilms}
Seo GH, Kim YM, Kim GH, Seo EJ, Choi JH, Lee BH, et~al.
\newblock Wilms tumor, aniridia, genitourinary anomalies, and mental retardation syndrome with deletion of chromosome 11p14. 3p12.
\newblock Journal of genetic medicine. 2018;15(1):38-42.

\bibitem{clericuzio2009diagnostic}
Clericuzio CL, Martin RA.
\newblock Diagnostic criteria and tumor screening for individuals with isolated hemihyperplasia.
\newblock Genetics in Medicine. 2009;11(3):220-2.

\end{thebibliography}

\clearpage
\section*{Appendix}
\setcounter{section}{0}
\renewcommand\thesection{\arabic{section}}

\section{Prompts}
\label{appendix:prompts}

The prompts for LLMs with and without RAG is given in Figure~\ref{fig:appendix_prompts}.

\begin{figure}[h]
\fcolorbox{black}{lightgray!15}{\parbox{.99\linewidth}
{
\textbf{Prompts for LLMs w/o RAG:} You are a helpful medical expert, and your task is to answer a multi-choice medical question. Please first choose the answer from the provided options and then provide the explanation.\textbackslash n

Question: \{question\}\textbackslash n

A. \{choices[0]\}\textbackslash n

B. \{choices[1]\}\textbackslash n

C. \{choices[2]\}\textbackslash n

D. \{choices[3]\}\textbackslash n

Answer: 
\\
\rule[0mm]{163mm}{0.15mm}
\\
\textbf{Prompts for RAG:} You are a helpful medical expert, and your task is to answer a multi-choice medical question using the relevant documents. Please first choose the answer from the provided options and then provide the explanation.\textbackslash n
\\
Relevant Documents:\textbackslash n
\\
\{Document[0]\}\textbackslash n
\\
\{Document[1]\}\textbackslash n
\\
$\cdots$
\\
\{Document[$k-1$]\}\textbackslash n
\\
Question: \{question\}\textbackslash n
\\
A. \{choices[0]\}\textbackslash n
\\
B. \{choices[1]\}\textbackslash n
\\
C. \{choices[2]\}\textbackslash n
\\
D. \{choices[3]\}\textbackslash n
\\
Answer: 
\vspace{2mm}
}}
\caption{Prompts for LLMs with and without RAG on the \Algnameabbr{} dataset.}
\label{fig:appendix_prompts}
\end{figure}

\section{\corpusname{} Complement Existing Textbooks, StatPearls, PubMed, and Wikipedia Corpus}
\label{appendix:pubmed_results}

We show \corpusname{}'s complement to existing Textbooks, StatPearls, PubMed, and Wikipedia corpus in Tables~\ref{tab:k5} and~\ref{tab:k7}.

\begin{table}[h]
    \centering
    \caption{Retrieved chunk number $k=5$ for each question.}
    \vspace{-3mm}
    \resizebox{\textwidth}{!}{
    \begin{tabular}{l|c|cccccc}
    \toprule
         Retriever & Corpus & Llama-2-7B & Mistral-7B-v0.2 & Phi-3-7B & Gemmma-1.1-7B & Qwen-2-7B & Average \\
    \midrule
        \multirow{8}{*}{MedCPT} & {Textbooks} & 43.1 & 57.1 & 72.9 & 56.7 & 66.8 & 59.3 \\
        & {Textbooks+ReCOP} & 44.6 & 59.9 & 75.1 & 59.6 & 69.9 & \textbf{61.8} \\
    \cmidrule[0.5pt]{2-8}
        & {StatPearls} & 41.0 & 55.3 & 72.2 & 56.5 & 64.9 & 58.0 \\
        & {StatPearls+ReCOP} & 44.2 & 59.0 & 74.4 & 60.4 & 68.6 & \textbf{61.3} \\
    \cmidrule[0.5pt]{2-8}
        & {PubMed} & 44.6 & 57.4 & 72.6 & 58.8 & 67.1 & 60.1 \\
        & {PubMed+ReCOP} & 45.6 & 58.8 & 74.6 & 62.1 & 69.2 & \textbf{62.1} \\
    \cmidrule[0.5pt]{2-8}
        & {Wikipedia} & 42.2 & 54.5 & 72.3 & 56.7 & 65.1 & 58.2 \\
        & {Wikipedia+ReCOP} & 45.4 & 58.5 & 74.6 & 60.5 & 68.3 & \textbf{61.5} \\
    \midrule
        \multirow{9}{*}{BM25} & {Textbooks} & 38.4 & 55.8 & 73.2 & 56.2 & 66.1 & 57.9 \\
        & {Textbooks+ReCOP} & 43.8 & 59.4 & 76.2 & 60.7 & 69.9 & \textbf{62.0} \\
    \cmidrule[0.5pt]{2-8}
        & {StatPearls} & 40.7 & 56.0 & 71.3 & 55.1 & 63.8 & 57.4 \\
        & {StatPearls+ReCOP} & 43.9 & 59.5 & 74.9 & 59.5 & 68.7 & \textbf{61.3} \\
    \cmidrule[0.5pt]{2-8}
        & {PubMed} & 44.4 & 58.6 & 72.3 & 57.6 & 68.7 & 60.3 \\
        & {PubMed+ReCOP} & 46.8 & 61.0 & 75.5 & 61.0 & 71.4 & \textbf{63.1} \\
    \cmidrule[0.5pt]{2-8}
        & {Wikipedia} & 38.0 & 56.4 & 71.3 & 53.8 & 64.9 & 56.9 \\
        & {Wikipedia+ReCOP} & 43.5 & 58.5 & 75.6 & 59.3 & 69.0 & \textbf{61.2} \\
    \bottomrule
    \end{tabular}
    }
    \label{tab:k5}
\end{table}

\begin{table}[h]
    \centering
    \caption{Retrieved chunk number $k=7$ for each question.}
    \vspace{-3mm}
    \resizebox{\textwidth}{!}{
    \begin{tabular}{l|c|cccccc}
    \toprule
         Retriever & Corpus & Llama-2-7B & Mistral-7B-v0.2 & Phi-3-7B & Gemmma-1.1-7B & Qwen-2-7B & Average \\
    \midrule
        \multirow{8}{*}{MedCPT} & {Textbooks} & 45.8 & 57.7 & 73.4 & 58.5 & 66.5 & 60.4 \\
        & {Textbooks+ReCOP} & 45.8 & 61.6 & 75.7 & 60.8 & 70.8 & \textbf{62.9} \\
    \cmidrule[0.5pt]{2-8}
        & {StatPearls} & 41.7 & 55.3 & 72.4 & 57.2 & 65.5 & 58.4 \\
        & {StatPearls+ReCOP} & 44.3 & 60.4 & 74.3 & 61.2 & 69.3 & \textbf{61.9} \\
    \cmidrule[0.5pt]{2-8}
        & {PubMed} & 45.7 & 58.2 & 72.4 & 60.0 & 69.2 & 61.1 \\
        & {PubMed+ReCOP} & 48.1 & 61.3 & 75.0 & 62.2 & 70.6 & \textbf{63.4} \\
    \cmidrule[0.5pt]{2-8}
        & {Wikipedia} & 41.8 & 54.4 & 72.5 & 56.6 & 64.7 & 58.0 \\
        & {Wikipedia+ReCOP} & 46.1 & 59.8 & 74.9 & 59.7 & 69.2 & \textbf{61.9} \\
    \midrule
        \multirow{9}{*}{BM25} & {Textbooks} & 40.1 & 56.9 & 72.8 & 57.1 & 67.1 & 58.8 \\
        & {Textbooks+ReCOP} & 43.8 & 61.0 & 75.5 & 62.0 & 70.2 & \textbf{62.5} \\
    \cmidrule[0.5pt]{2-8}
        & {StatPearls} & 40.7 & 56.8 & 71.6 & 55.8 & 65.4 & 58.1 \\
        & {StatPearls+ReCOP} & 45.0 & 61.8 & 75.1 & 60.3 & 69.5 & \textbf{62.3} \\
    \cmidrule[0.5pt]{2-8}
        & {PubMed} & 44.0 & 58.2 & 73.3 & 57.9 & 68.4 & 60.4 \\
        & {PubMed+ReCOP} & 46.5 & 62.1 & 75.1 & 62.5 & 72.0 & \textbf{63.6} \\
    \cmidrule[0.5pt]{2-8}
        & {Wikipedia} & 39.5 & 58.2 & 73.0 & 54.0 & 66.5 & 58.2 \\
        & {Wikipedia+ReCOP} & 45.6 & 61.9 & 76.5 & 60.3 & 69.7 & \textbf{62.8} \\
    \bottomrule
    \end{tabular}
    }
    \label{tab:k7}
\end{table}

\section{Explanations of LLMs with \corpusname{}}
\label{appendix:explanation}

We give more case studies of natural language explanations in Figures~\ref{fig:exp_case_study2}, \ref{fig:exp_case_study3}, \ref{fig:exp_case_study4}, and \ref{fig:exp_case_study5}.

\begin{figure}[h]
\fcolorbox{black}{lightgray!15}{\parbox{.99\linewidth}
{
\textbf{Question:} Primary site of infection in Cryptococcosis is 
\\
\textbf{Choices:}
(A) Adrenal glands (B) Bone (C) Central nervous system (D) Lungs 
\vspace{2mm}
\\
\textbf{LLMs w/o RAG Answer:} \!\!\!\! \colorbox{lightred}{C. Central nervous system.} \textbf{Explanation:} Cryptococcosis is a fungal infection caused by the yeast Cryptococcus neoformans. \colorbox{lightred}{The primary site of infection is typically the central nervous system (CNS)}, where the infection can cause meningitis or encephalitis. The disease can also spread to other parts of the body, including the lungs, but the CNS is the most common site of infection and the one that can cause the most serious health problems. Therefore, the answer is C. Central nervous system.
\vspace{2mm}
\\
\textbf{\corpusname{}-RAG Answer:} \!\!\!\! \colorbox{lightgreen}{D. Lungs.} \textbf{Explanation:} \!\!\!\! \colorbox{lightgreen}{The primary site of infection in Cryptococcosis is the lungs. The} \colorbox{lightgreen}{infection may remain localized to the lungs or spread to other parts of the body, including the central nervous} \colorbox{lightgreen}{system, bones, and skin.} However, the lungs are the most common site of infection.
\vspace{2mm}
\\
\textbf{Golden Answer:} \!\!\!\! \colorbox{lightgreen}{D. Lungs.} 
\textbf{Explanation from Literature:} \!\!\!\! \colorbox{lightgreen}{For the Cryptococcosis pathogen, the primary site of} \colorbox{lightgreen}{entry is the lungs. Clinical symptoms can range from asymptomatic colonization in the lungs to widespread}  \colorbox{lightgreen}{systemic infection. \cite{carrasco2016cutaneous}}
}}
\caption{Natural language explanations to the question of Cryptococcosis.}
\label{fig:exp_case_study2}
\end{figure}

\begin{figure}[h]
\fcolorbox{black}{lightgray!15}{\parbox{.99\linewidth}
{
\textbf{Question:} Achondroplasia shows which of the following type of inheritance
\\
\textbf{Choices:}
(A) Autosomal Dominant (B) Autosomal Recessive (C) X-Linked Dominant (D) X-Linked Recessive 
\vspace{2mm}
\\
\textbf{LLMs w/o RAG Answer:} \!\!\!\!\colorbox{lightred}{B. Autosomal Recessive.} \textbf{Explanation:} Achondroplasia is a genetic disorder caused by mutations in the FGFR3 gene, which is located on the short arm of chromosome 4 (autosome). \colorbox{lightred}{It is an autosomal recessive disorder, meaning that an individual must inherit two copies of the mutated gene,} \colorbox{lightred}{one from each parent, to develop the condition.} Autosomal dominant disorders, on the other hand, only require one mutated gene to be expressed, while X-linked dominant and X-linked recessive disorders are related to genes on the X chromosome.
\vspace{2mm}
\\
\textbf{\corpusname{}-RAG Answer:} \!\!\colorbox{lightgreen}{A. Autosomal Dominant.} \textbf{Explanation:}\colorbox{lightgreen}{Achondroplasia is an autosomal dominant disorder}
\colorbox{lightgreen}{, which means that a single copy of the mutated gene in each cell is sufficient to cause the disorder.} The risk of passing the abnormal gene from an affected parent to an offspring is 50\% for each pregnancy.
\vspace{2mm}
\\
\textbf{Golden Answer:} \!\!\!\! \colorbox{lightgreen}{A. Autosomal Dominant.} \textbf{Explanation from Literature: } \!\!\!\! Achondroplasia, the most prevalent form of dwarfism in humans, occurs in approximately 1 in 15,000 to 25,000 births, with 80\% of cases arising sporadically. \colorbox{lightgreen}{Achondroplasia is characterized as an autosomal dominant genetic disorder with full penetrance  \cite{ornitz2017achondroplasia}}
}}
\caption{Natural language explanations to the question of Achondroplasia.}
\label{fig:exp_case_study3}
\end{figure}

\begin{figure}[h]
\fcolorbox{black}{lightgray!15}{\parbox{.99\linewidth}
{
\textbf{Question:} Which drug is used for Achalasia cardia
\\
\textbf{Choices:}
(A) Nifedipine (B) Propranolal (C) Atenolol (D) Bethnecol 
\vspace{2mm}
\\
\textbf{LLMs w/o RAG Answer:} \!\!\!\! \colorbox{lightred}{D. Bethanechol.} \textbf{Explanation:} \colorbox{lightred}{Achalasia is a condition where the muscles in the lower} \colorbox{lightred}{ esophagus fail to relax and allow food to pass into the stomach.} \colorbox{lightred}{Bethanechol is a parasympathomimetic drug that } \colorbox{lightred}{increases the activity of the parasympathetic nervous system, which helps to relax the muscles in the lower } \colorbox{lightred}{esophagus and improve the passage of food.} Nifedipine is a calcium channel blocker used to treat hypertension and angina, Propranolol and Atenolol are both beta-blockers used to treat various conditions including hypertension, angina, and heart rhythm disorders.
\vspace{2mm}
\\
\textbf{\corpusname{}-RAG Answer:} \!\!\!\! \colorbox{lightgreen}{A. Nifedipine.} \textbf{Explanation:} \!\!\!\! \colorbox{lightgreen}{Nifedipine is a calcium channel blocker that is used in the} \colorbox{lightgreen}{treatment of achalasia.} It helps to relax the lower esophageal sphincter muscle, making it easier for food to pass through. The other options, Propranolol and Atenolol, are beta-blockers, which are not used in the treatment of achalasia. Bethanechol is a parasympathomimetic drug that is used to increase the contractions of the gastrointestinal tract, and it is not typically used in the treatment of achalasia.
\vspace{2mm}
\\
\textbf{Golden Answer:} 
\!\!\!\! \colorbox{lightgreen}{A. Nifedipine.} \textbf{Explanation from Literature:} \!\!\!\! \colorbox{lightgreen}{Sublingual administration of nifedipine is highly } \colorbox{lightgreen}{effectivefor treating achalasia for two reasons: 1) it lowers the pressure of the lower esophageal sphincter (LESP),} \colorbox{lightgreen}{and 2) it bypasses the esophagogastric junction, which has abnormally high pressure in achalasia patients.
 \cite{hongo1982clinical}}
}}
\caption{Natural language explanations to the question of  Achalasia cardia.}
\label{fig:exp_case_study4}
\end{figure}

\begin{figure}[h]
\fcolorbox{black}{lightgray!15}{\parbox{.99\linewidth}
{
\textbf{Question:} In which of the following conditions, Aniridia and Hemihyperophy are most likely present?
\\
\textbf{Choices:}
(A) Neuroblastoma (B) Wilms' tumour (C) Non-Hodgkin's lymphoma (D) Germ cell tumour 
\vspace{2mm}
\\
\textbf{LLMs w/o RAG Answer:} \!\!\!\! \colorbox{lightred}{None of the given conditions.} \textbf{Explanation:}  \colorbox{lightred}{Neither of the given conditions, Aniridia}  \colorbox{lightred}{ and Hemihyperphoria, are related to the conditions listed in the choices.} Aniridia is a condition characterized by the absence of the iris in the eye, while Hemihyperphoria is a condition where one eye is slightly higher than the other in the orbit. Therefore, none of the given conditions (Neuroblastoma, Wilms' tumour, Non-Hodgkin's lymphoma, and Germ cell tumour) are most likely present when Aniridia and Hemihyperphoria are present.
\vspace{2mm}
\\
\textbf{\corpusname{}-RAG Answer:} \!\!\!\! \colorbox{lightgreen}{B. Wilms' tumour.} \textbf{Explanation:} \!\!\!\! \colorbox{lightgreen}{Aniridia is a condition characterized by the partial} \colorbox{lightgreen}{orcomplete absence of the iris in the eye. It can occur as an isolated condition or in association with other} \colorbox{lightgreen}{conditions such as Wilms' tumor, which is a type of kidney tumor.} \colorbox{lightgreen}{Hemihyperphosphatasia is characterized by the} \colorbox{lightgreen}{ excessive production of alkaline phosphatase in one side of the body.} Therefore, the correct answer is B. Wilms' tumour, as it is the condition most likely to be associated with both Aniridia and Wilms' tumor.
\vspace{2mm}
\\
\textbf{Golden Answer:} \!\!\!\! \colorbox{lightgreen}{B. Wilms' tumour.} \textbf{Explanation from Literature:} \!\!\!\! \colorbox{lightgreen}{Wilms tumor, aniridia, genitourinary}  \colorbox{lightgreen}{abnormalities, and intellectual disability always appear together in the Wilms' tumor, aniridia, genitourinary} \colorbox{lightgreen}{anomalies, and mental retardation syndrome.\cite{seo2018wilms}}

\colorbox{lightgreen}{Isolated hemihyperplasia is linked to a higher risk of developing embryonal tumors, particularly Wilms tumor and}
\colorbox{lightgreen}{hepatoblastoma\cite{clericuzio2009diagnostic}}
}}
\caption{Natural language explanations to the question of Aniridia and Hemihyperophy.}
\label{fig:exp_case_study5}
\end{figure}

\section{Algorithm of Combining \corpusname{} with Other Corpus}
\label{appendix:comb_alg}

We give the algorithm of combining \corpusname{} with baseline corpus in Algorithm~\ref{alg:eb_mc_rag}.

\begin{algorithm}
\caption{Entropy-Aware Multi-Corpora Retrieval Augmented Generation}\label{alg:eb_mc_rag}
\begin{algorithmic}
\State \textbf{Input:} Query $q$
\State \textbf{Corpora:} Corpus X $C_X$, Corpus Y $C_Y$
\State \textbf{Model:} Large Language Model $\mathcal{M}$
\State \textbf{Retrievers:} Retriever for Corpus X $\mathcal{R}_X$, Retriever for Corpus Y $\mathcal{R}_Y$
\State \textbf{Parameters:} Number of relevant documents $k$
\State \textbf{Output:} Answer $a \in \{\text{A, B, C, D}\}$

\State

\State \textbf{Step 1: Retrieve Relevant Documents}
\State $D_X \gets \text{RetrieveRelevantDocs}(q, C_X, \mathcal{R}_X, k)$
\State $D_Y \gets \text{RetrieveRelevantDocs}(q, C_Y, \mathcal{R}_Y, k)$

\State

\State \textbf{Step 2: Generate Response with LLM}
\State $P_X \gets \text{GenerateResponse}(q, D_X, \mathcal{M})$
\State $P_Y \gets \text{GenerateResponse}(q, D_Y, \mathcal{M})$

\State

\State \textbf{Step 3: Compute Probabilities for Options}
\State $\text{Probs}_X \gets \text{ComputeProbs}(P_X, \{\text{A, B, C, D}\})$
\State $\text{Probs}_Y \gets \text{ComputeProbs}(P_Y, \{\text{A, B, C, D}\})$

\State

\State \textbf{Step 4: Calculate Entropy}
\State $H_X \gets \text{CalculateEntropy}(\text{Probs}_X)$
\State $H_Y \gets \text{CalculateEntropy}(\text{Probs}_Y)$

\State

\State \textbf{Step 5: Select Answer Based on Entropy}
\If {$H_X < H_Y$}
    \State $a \gets \text{argmax}(\text{Probs}_X)$
\Else
    \State $a \gets \text{argmax}(\text{Probs}_Y)$
\EndIf

\State

\State \textbf{Return} $a$
\end{algorithmic}
\end{algorithm}

\end{document}